%
%
\magnification=1200
\baselineskip=15pt
\def\F{{\cal F}}
\def\d{{\rm d}}
\def\G{\Gamma}
\def\L{\bar{\Lambda}}
\def\l{\,{\rm log\,}}
\def\al{\bar a_l}
\def\ak{\bar a_k}

\def\xk{x_k^-}
\def\x1{x_1^-}
\def\dko{\delta_k^{(1)}}
\def\dkt{\delta_k^{(2)}}
\def\dkT{\delta_k^{(3)}}
\def\dkf{\delta_k^{(4)}}
\def\dlo{\delta_l^{(1)}}
\def\dlt{\delta_l^{(2)}}
\def\dlT{\delta_l^{(3)}}
\def\dlf{\delta_l^{(4)}}
\def\dmo{\delta_m^{(1)}}
\def\dmt{\delta_m^{(2)}}

\def\12{{1\over 2}}

\rightline{UCLA/96/TEP/28}
\bigskip
\centerline{\bf THE EFFECTIVE PREPOTENTIAL OF N=2 SUPERSYMMETRIC}
\bigskip
\centerline{{\bf SU($N_c$) GAUGE THEORIES}\footnote*{Research
supported in part by the National Science Foundation
under grants PHY-95-31023 and DMS-95-05399}}
\bigskip
\bigskip
\centerline{{\bf Eric D'Hoker}${}^1$, 
												{\bf I.M. Krichever}${}^2$,
            {\bf and D.H. Phong} ${}^3$}
\bigskip
\centerline{${}^1$ Theory Division CERN}
\centerline{1211, Geneva 23, Switzerland}
\centerline{and}
\centerline{Department of Physics}
\centerline{University of California, Los Angeles, CA 90024, USA}
\centerline{e-mail: dhoker@phys.ucla.edu}
\bigskip 
\centerline{${}^2$ Landau Institute for Theoretical Physics}
\centerline{Moscow 117940, Russia}
\centerline{and} 
\centerline{Department of Mathematics}
\centerline{Columbia University, New York, NY 10027, USA}
\centerline{e-mail: krichev@math.columbia.edu}
\bigskip
\centerline{${}^3$ Department of Mathematics}
\centerline{Columbia University, New York, NY 10027, USA}
\centerline{e-mail: phong@math.columbia.edu}

\vfill\break
\centerline{\bf ABSTRACT}
\bigskip
We determine the effective prepotential for N=2 supersymmetric
SU($N_c$) gauge theories with an arbitrary number of flavors $N_f<2N_c$,
from the exact solution constructed out of spectral curves.
The prepotential is the same for the several models of spectral curves
proposed in the literature. It has to all orders the logarithmic
singularities of the one-loop perturbative corrections, thus
confirming the non-renormalization theorems from supersymmetry.
In particular, the renormalized order parameters and their
duals have all the correct monodromy transformations
prescribed at weak coupling.
We evaluate explicitly the contributions of one- and two-instanton
processes.

\vfill\break 

\centerline{\bf I. INTRODUCTION}

\bigskip

A recurrent feature in many phenomenologically interesting
supersymmetric field as well as string theories is the
presence of a flat potential, resulting in a
moduli space of inequivalent vacua. For N=2 supersymmetric
SU(2) gauge theories, Seiberg and Witten [1] have shown
in their 1994 seminal work how to extract the physics
from a fibration, over the space of
vacua, of {\it spectral curves}, i.e., Riemann surfaces equipped with a 
meromorphic 1-form $\d\lambda$,
whose periods determine the spectrum of BPS states.
Rapid progress has led since then to
an almost complete classification [2] of the spectral curves
for N=2 supersymmetric theories with arbitrary
gauge groups and number of quark hypermultiplets $N_f$,
and in particular for SU($N_c$) theories with $N_f<2N_c$
[3][4]. Although these curves are 
strongly suggested, e.g.,
by consistency requirements in various limits
(such as the classical limit or the infinite mass limit as a quark
decouples), or by analogies with
singularity theory and soliton theory, they
are often still conjectural. In particular, in most
cases, even a direct verification that the spectral curves
do reproduce the known perturbative correction [5] to the prepotential
is lacking. Nor has the predictive power of the spectral curves
been put to effective use, for example in
deriving the contributions to the prepotential
of $d$-instanton processes.

\medskip

Mathematically, the evaluation of the prepotential presents
the unusual feature that the $B$-periods of the meromorphic form
$\d\lambda$ have to be expressed in terms of its $A$-periods,
rather than in terms of the moduli parameters which occur
explicitly as coefficients in the defining equation
of the Riemann surface. It is well-known that, as functions
of moduli parameters, the vector of both $A$ and $B$-periods
satisfies Picard-Fuchs differential equations. Picard-Fuchs
equations have been applied successfully in the case
of SU(3) by Klemm et al. and Ito and Yang [6], but their complexity
increases rapidly with the number of
colors, and makes it difficult to treat the case of general SU($N_c$),
even without hypermultiplets.

\medskip

In this paper, we develop methods for determining the
prepotential from the spectral curves
for an arbitrary number of colors and flavors in the $N_f<2N_c$
asymptotically free case, in the regime
where the dynamically generated scale $\Lambda$ of the theory 
is small. We show that, in this regime, the full expansion of 
the $A$-periods
can actually been calculated by the method of residues.
For the $B$-periods, we provide a simple algorithm
to arbitrary order of multi-instanton processes. This
method is based on an analytic continuation in an
auxiliary parameter $\xi$, which allows us to
expand the form $\d\lambda$ into a series of
rational functions that can thus be integrated in closed form.
The method is powerful enough to let us identify completely
all the logarithmic singularities of the prepotential
and, in particular, to confirm the non-renormalization theorems
resulting from supersymmetry.

\medskip

We have worked out explicitly the perturbative corrections,
as well as the contributions of up to 2-instanton processes,
to the prepotential for any gauge group SU($N_c$)
and arbitrary $N_f<2N_c$. The perturbative part does
coincide with the one calculated from one-loop
effects in field theory [5]. The instanton contributions
are in complete agreement with 
all known expressions, and in particular, with
(a) the 1-instanton contribution for
pure (with $N_f=0$) SU(3) obtained in [6]
via Picard-Fuchs equations; (b) more
generally, with the 1-instanton contribution for
pure SU($N_c$) obtained in [7]
via holomorphicity arguments; and (c), with
the 2-instanton contribution for SU(2) with $N_f<4$
hypermultiplets obtained recently in [8] from first
principles.

\medskip

We note that, although the spectral
curves proposed in [3] and [4] differ when $N_f\geq N_c+2$, the
corresponding prepotentials are the same. 

\medskip

We observe that the final form
of the prepotential is very simple, and results from a large number
of remarkable cancellations. This suggests the presence
of a deeper geometric structure on the moduli space of
vacua. Some of it has recently come to light,
especially in connection with 
Calabi-Yau compactifications in string theory [9], 
topological field theories and WDVV equations [10],
symplectic geometry 
[11][4]
and integrable models [12][13] (see especially [13],
where a closed form for the prepotential is written
in terms of parameters inherent to Whitham-averaged
hierarchies). It seems that further investigation along
these lines is warranted.

\bigskip

\centerline{\bf II. INTEGRABILITY CONDITIONS AND MODEL INDEPENDENCE}

\bigskip

We consider N=2 supersymmetric SU($N_c$) gauge theories with $N_f$ quark
flavors, $N_f<2N_c$. The field content is an N=2 chiral multiplet and
$N_f$ hypermultiplets of bare masses $m_j$. The N=2 chiral
multiplet contains a complex scalar field $\phi$ in the adjoint
representation. The flat directions in the potential correspond to
$[\phi,\phi^{\dagger}]=0$, so that the classical moduli space
of vacua is $N_c-1$ dimensional, and can be parametrized by the
eigenvalues 
$$\ak,\ \ 1\leq k\leq N_c,\ \ \sum_{k=1}^{N_c}\ak=0$$ 
of $\phi$.
For generic $\ak$, the SU($N_c$) gauge symmetry is broken
down to ${\rm U(1)}^{N_c-1}$. In the N=1 formalism,
the Wilson effective Lagrangian of the quantum theory  to leading
order in the low momentum expansion is 
of the form
$$
{\cal L}={\rm Im}{1\over 4\pi}[\int \d^4\theta 
{\partial\F(A) \over \partial A^i} \overline{A^i}+
\12\int\d^2\theta
{\partial ^2 \F(A) \over\partial A^i\partial A^j} W^iW^j]
$$
where the $A^i$'s are N=1 chiral superfields whose scalar 
components correspond to 
the $\bar a_i$'s, and $\F$ is the holomorphic prepotential. 
For SU($N_c$) gauge theories
with $N_f<2N_c$ flavors, general arguments, based on 
holomorphicity of $\F$,
perturbative non-renormalization beyond 1-loop order, 
the nature of instanton
corrections and the restrictions of $U(1)_R$ invariance, 
suggest that $\F$
should have the following form 
$$
\eqalignno{ 
\F(A)=&{1\over 2\pi i}(2N_c-N_f)\sum_{i=1}^{N_c}A_i^2\cr
&-{1\over 8\pi i}\big(\sum_{k,l=1}^{N_c}(A_k-A_l)^2
\l{(A_k-A_l)^2\over\Lambda^2}
-\sum_{k=1}^{N_c}\sum_{j=1}^{N_f}(A_k+m_j)^2\l 
{(A_k+m_j)^2\over\Lambda^2}\big)\cr
&+\sum_{d=1}^{\infty}\F_d (A_k) \Lambda^{(2N_c-N_f)d}
&(2.1)\cr}
$$
The terms on the right hand side are respectively the 
classical prepotential, the contribution of
perturbative one-loop effects (higher loops do not
contribute in view of non-renormalization
theorems), and the contributions of $d$-instantons processes.

\medskip

\noindent{\bf a) The spectral curves of the effective theory}

The Seiberg-Witten Ansatz for determining the full 
prepotential $\F$
(as well as the spectrum of BPS states) is based on the 
choice of a fibration of
spectral curves over the space of vacua, and of a meromorphic 
1-form $\d\lambda$
over each of these curves. The renormalized order parameters 
$a_k$'s
of the theory, their duals $a_{D,k}$'s, and the prepotential
$\F$ are then given by
$$
2\pi i \,a_k= \oint_{A_k}\d\lambda,\qquad 
2 \pi i \,a_{D,k}= \oint_{B_k}\d\lambda, \qquad
a_{D,k}={\partial\F\over\partial a_k}
\eqno(2.2)
$$   
with $A_k$, $B_k$ a canonical basis of homology cycles on the 
spectral curves.
\medskip
For SU($N_c$) theories with $N_f$ hypermultiplets, the following 
candidate spectral curves ([3][4]) and meromorphic forms $\d\lambda$ 
have been proposed
$$
\eqalignno{y^2&=A^2(x)-B(x)\cr
\d\lambda&={x\over y}\big(A'-\12 (A-y){B'\over B}\big)\d x
&(2.3)\cr}
$$
where $A(x)$ and $B(x)$ are polynomials in $x$ of respective 
degrees $N_c$
and $N_f$, whose coefficients vary with the physical
parameters of the theory. More specifically, let   
$\Lambda$ be the dynamically generated scale of the theory,
$\bar s_i$, $0\leq i\leq N_c$, and $t_p(m)$, $1\leq p\leq N_f$, 
be
the $i$-th and $p$-th symmetric polynomials in $\ak$ and $m_j$
$$
\bar s_i=(-1)^i\sum_{k_1<\cdots<k_i}\bar a_{k_1}\cdots\bar a_{k_i},\ 
\ \ \ t_p(m)=\sum_{j_1<\cdots<j_p}m_{j_1}\cdots m_{j_p},
\eqno(2.4)
$$
and let $\bar\sigma_i$, $0\leq i\leq N_c$, and $\tilde t_q$, be
defined by $\bar\sigma_0=1$, $\bar\sigma_1=0$, and 
$$
\eqalignno{\delta_{p,0}&=\sum_{i+j=p}\bar s_i\bar\sigma_j,
\ 0\leq p\leq N_c\cr
\tilde t_q&=\sum_{p+i=q}t_p(m)\bar\sigma_i 
&(2.5)\cr}
$$
Then the polynomials $A(x)$ and $B(x)$ are given by
$$
\eqalignno{
A(x)&=C(x)+{\Lambda^{2N_c-N_f}\over 4}T(x)\cr
B(x)&=\Lambda^{2N_c-N_f}\prod_{j=1}^{N_f}(x+m_j)
&(2.6)\cr}
$$
where 
$$
C(x)=\prod_{k=1}^{N_c}(x-\ak)=x^{N_c}+\sum_{i=2}^{N_c}\bar s_ix^{N_c-i}
\eqno(2.7)
$$
and $T(x)$ is a polynomial of degree $N_f-N_c$ when  
$N_f-N_c \geq 0$, and is 
zero when $N_f-N_c < 0$.
$$
T(x)=\sum_{p=0}^{N_f-N_c}t_px^{N_f-N_c-p},\
{\rm or}\ \ \ \ T(x)=\sum_{p=0}^{N_f-N_c}\tilde t_px^{N_f-N_c-p},
\eqno(2.8)
$$
depending on whether we consider the model
of [3] or of [4] respectively. Notice that $T(x)$ is independent 
of $\bar s_i$ for
the model of [3], but does depend on $\bar s_i$ for the model
of [4]. Evidently, the two choices for
$T(x)$ agree when $N_f\leq N_c+1$.

\medskip

It has become customary to describe the moduli space of supersymmetric
vacua in terms of SU($N_c$) invariant polynomials in $\bar a_k$, such as 
the symmetric polynomials $\bar s_i$ or $\bar \sigma _i$ of (2.4)
and (2.5). In particular, the approaches to calculating the prepotential
using the Picard-Fuchs equations [6] as well as the calculations of 
strong coupling monodromies make use of such variables. Within the
context of our own calculations, which are performed for arbitrary
$N_c$ and $N_f < 2 N_c$, the use of the polynomials $\bar s_i$ and 
$\bar \sigma _i$
does not appear to be as fruitful. Instead, we shall parametrize 
supersymmetric vacua
by $\ak$ at the classical level and by $a_k$ at the quantum level. 
In doing so, we keep in
mind that 
$$\sum _k \ak = \sum _k a_k =0,$$ 
and that the space of all such $\ak$ must be
coseted out by the permutation group (i.e. the Weyl group of SU($N_c$)). 
Our final result for the prepotential will be expressed in terms of 
invariant functions as well,
but their translation into $\bar s_i$ variables is cumbersome and 
unnecessary. This is
perhaps not so surprising, since even the one-loop answer is not 
simply rewritten in
terms of the functions $\bar s_i$.

\medskip

\noindent{\bf b) Integrability conditions for the prepotential}

The key features of the meromorphic 1-form $\d\lambda$ are the
following

\noindent{(i)} The residue of $\d\lambda$ at the pole $x=-m_j$ (and
the lower sheet of the Riemann surface) is clearly
equal to $-m_j$, as required by the linearity of the BPS mass formula 
in terms
of the quark masses.

\medskip

\noindent{(ii)} The derivatives of $\d\lambda$ with respect to the moduli
parameters $\ak$ are holomorphic. Actually, as discussed in [4], there are
two natural connections with which
$\d\lambda$ can be differentiated along the $\ak$ directions. With respect to 
the first, $\nabla^{(E)}$,
the values of the abelian integral $\l(y+A(x))$ are kept fixed, and
$\nabla^{(E)}\d\lambda$ is the holomorphic 1-form
$$
\nabla^{(E)}(\d\lambda)=-{1\over y}\big(\delta A-\12{\delta B\over B}(A-y)
\big)\d x
\eqno(2.9)
$$
With respect to the second, $\nabla^{(x)}$, the values of $x$ are kept fixed,
and $\nabla^{(x)}\d\lambda$ differs from the above holomorphic form
by the differential of a meromorphic function
$$\nabla^{(x)}\d\lambda
=-{1\over y}\big(\delta A-\12{\delta B\over B}(A-y)\big)\d x
+\d\big(-{x\over y}(\delta A-\12{\delta B\over B}(A-y))\big)
\eqno(2.10)
$$ 
Since exact differentials do not contribute to the periods (2.2), 
we can ignore
this ambiguity and adopt the right hand side of (2.9) for the 
derivatives of 
$\d\lambda$. We also note
that we may simplify (2.9) to 
$-{1\over y}(\delta A-\12{\delta B\over B}A)$,
since the term ${\delta B\over B}A$ is $y$-independent, 
and does not contribute
to the periods. Similarly, the corresponding term $x{B'\over B}A$
in (2.3) may be ignored.
\medskip
It follows from (ii) that the derivatives $\nabla_{\ak}\d\lambda$
of $\d\lambda$ with respect to
$\ak$, say for $k=2,\cdots, N_c$, form a basis $\omega_k$ of holomorphic 
1-forms
on the spectral curve, at least for $\Lambda$ small (This is
evident if we differentiate with respect to the
$\bar s_i$'s. The derivatives with
respect to the $\bar a_k$ amount then to
a change of basis). An important 
observation is that this implies
that the dual periods $a_{D,k}$, given by integrals over the
$B_k$ cycles, do form a gradient, i.e., that the Ansatz (2.2) does guarantee
the existence of a prepotential. In fact, for each fixed $\Lambda$,
we can view $(\ak)\rightarrow (a_k)$ as a change of variables. Thus
the dual variables $a_{D,l}$, which depend originally on $\ak$, $m_j$, and
$\Lambda$, can be thought of as functions $a_{D,l}(a,m,\Lambda)$ of
$a_k$, $m_j$, and $\Lambda$. By the chain rule,
$$
{\partial a_{D,l}\over\partial a_m}=\sum_{k=2}^{N_c}
{\partial a_{D,l}\over\partial\ak}{\partial\ak\over\partial a_m}
=\sum_{k=2}^{N_c}
\big({\partial a_{D}\over\partial\bar a}\big)_{lk}
\big({\partial a\over\partial \bar a}\big)_{km}^{-1}
$$
Since $\big({\partial a_{D}\over\partial\bar a}\big)_{lk}$ and
$\big({\partial a\over\partial \bar a}\big)_{lk}$ are respectively
the matrices of periods of the Abelian differentials $\omega_k$ over
the cycles $B_l$ and $A_l$ of the canonical homology basis,
we recognize the matrix ratio above as just the period
matrix $\tau_{lm}$ of the spectral curve. In view of the symmetry
of period matrices, we have then established the integrability
condition
$$
{\partial a_{D,l}\over\partial a_m}=\tau_{lm}=\tau_{ml}
={\partial a_{D,m}\over\partial a_l}
\eqno(2.11)
$$
and thus the existence of a prepotential function $\F$.

\medskip

\noindent{\bf c) Model independence of the prepotential}

The functional form of the renormalized order parameters 
$a_k$ and their duals $a_{D,k}$ as functions of the classical 
parameters $\ak$ is different for the two models listed in (2.8).
Both $a_k$ and $a_{D,k}$, as functions of $\ak$, depend
non-trivially upon the parameters $t_p$ or $\tilde t_p$ that
specify the function $T(x)$. This dependence will be 
exhibited explicitly in sect. III.a, where an exact expression
for $a_k (\ak, t_p, m_j ;\L) $ will be given.
We shall now establish that the prepotential $\F$,
as well as the renormalized dual order parameters $a_{D,k}$, expressed as
functions of the renormalized order parameters $a_k$, are 
{\it independent} of $t_p$ or $\tilde t_p$.
Thus, the prepotential $\F$, expressed in terms of $a_k$
is independent of the models in (2.8), and both 
models yield the same prepotential $\F$, and thus the 
same low energy physics.

To establish independence of $T(x)$, we notice that all
$\ak$-dependence of $\d\lambda$, and thus of $a_k$ and 
$a_{D,k}$, resides in the function $A(x)$ of (2.3)
and (2.6). Also, $A(x)$ is the only place where the
dependence on $T(x)$ enters. Now, all dependence on
$T(x)$ may be absorbed in a redefinition of the classical
order parameters $\ak$, since the addition of $T(x)$
just modifies the bare parameters $\bar s_i$ in (2.7) to 
parameters $\tilde s_i$, as follows
$$
\eqalign{
A(x) = & ~x^{N_c} + \sum _{i=2} ^{N_c} \tilde s_i x^{N_c -i}
=\prod _{k=1} ^{N_c} (x- \tilde a_k) \cr
\tilde s_i = &\bar s_i + { 1\over 4} \Lambda ^{2N_c -N_f} t_i 
\cr}
\eqno (2.12)
$$
Thus, the addition of $T(x)$ may be absorbed by defining new
classical order parameters $\tilde a_k$, as shown above, and we have
$$
\eqalign{
a_k = & ~ a_k (\al, t_p, m_j; \Lambda) 
= a_k ( \tilde a_l, 0, m_j; \Lambda ) \cr
a_{D,k} = & ~ a_{D,k} (\al, t_p, m_j; \Lambda) 
= a_{D,k} ( \tilde a_l, 0, m_j; \Lambda )
\cr}
$$ Eliminating $\tilde a_l$ between $a_k$ and $a_{D,k}$ or $\F$
precisely amounts to eliminating $\al$ when $T(x)=0$, which
demonstrates that $a_{D,k}(a_l)$ and $\F (a_l)$ are independent
of $T(x)$.

\medskip

In view of this model independence of $\F$, we may set $\tilde a_k=\ak$
and $T=0$ when calculating $\F$, or when calculating
$a_{D,k}$ as functions of $a_k$. 

\bigskip

\centerline{\bf III. LOGARITHMS AND NON-RENORMALIZATION}

\bigskip

In this section, we shall describe an algorithm for calculating
the renormalized order parameters $a_k$ and their duals
$a_{D,k}$ to any order of perturbation theory, in
the regime where $\Lambda$ is small, and the variables
$a_k$ (equivalently, their classical counterparts
$\ak$) are well-separated. In particular,
we establish a general non-renormalization theorem of
the prepotential $\F$, expressed in terms of the renormalized 
$a_k$ : $\F$ contains only those logarithmic terms appearing 
in the expression (2.1).

\medskip

We begin by describing more precisely the representation
of the Riemann surface (2.3) as a double cover of the complex plane,
and our choice of homology cycles. It is convenient to
set
$$
\L=\Lambda^{N_c-N_f/ 2}
\eqno(3.1)
$$
The branch points $x_k^{\pm}$, 
$1\leq k\leq N_c$, of the surface are defined then by
$$A(x_k^{\pm}) ^2 -B(x_k^{\pm})=0.$$ 
For $\L$ small, $x_k ^\pm$ are
just perturbations of the $\ak$'s. 
We view the Riemann surface as two copies of the complex plane,
cut and joined along slits from $x_k^-$ to $x_k^+$. 
A canonical homology basis
$A_k,B_k$, $2\leq k\leq N_c$, is obtained by choosing $A_k$ to
be a simple contour enclosing the slit from $x_k^-$ to $x_k^+$, 
and $B_k$
to consist of the curves going from $x_k^-$ to $x_1^-$ on each sheet.
With this choice, we do have
$\#(A_j\cap A_k)=\#(B_j\cap B_k)=0$, $\#(A_j\cap B_k)=\delta_{jk}$.

\medskip 

\noindent  {\bf a) Expansion of the order parameters $a_k$}

First, we recall that the renormalized order parameters $a_k$ are given by combining
(2.2) and (2.3)
$$
2\pi i ~a_k=\oint_{A_k}\d\lambda=\oint_{A_k}\d x
{x({A'\over A}-\12 {B'\over B})\over\sqrt{1-{B\over A^2}}}
\eqno(3.2) 
$$
We can now reduce the evaluation of this integral to a set of residue calculations in 
the following way. We fix the location of the contour $A_k$, such that its distance
away from the branch cut between $x_k^-$ and $x_k^+$ is much 
larger than $\L$. This can
always be achieved since we assumed that the points $\ak$ are well separated. Having 
fixed $A_k$, we consider $a_k$ in a power series in $\L ^2$, i.e. 
this allows us to
expand the denominator in a convergent power series in $B(x)$ 
$$
2\pi i~a_k=\oint_{A_k}\d x\, x{A'\over A}+\sum_{m=1}^{\infty}
{\G(m+\12)\over\G(\12)\G(m+1)}\oint_{A_k}\d x x({A'\over A}-\12 {B'\over B})
({B\over
A^2})^m
\eqno(3.3)
$$
Similarly, powers and derivatives of $A(x)$ may be expanded in a series in powers of 
$\L ^2 T(x)$, leaving integrals of rational functions in $x$, which can be evaluated
using residue methods. It is very convenient to introduce the residue 
functions
$R_k(x)$ and $S_k(x)$, defined by
$$
{B(x) \over C(x)^2} = {\L ^2 \over (x- \ak )^2} S_k (x)
\eqno(3.4)
$$
and
$$
{T(x) \over C(x)} = {1 \over x- \ak } R_k (x)
\eqno(3.5)
$$
The first term on the right hand side of (3.3) is evaluated by expressing
${A'\over A}$ as
$${A'(x)\over A(x)}={C'(x)\over C(x)}+{\d\over\d x}\l(1+\L^2{T(x)\over C(x)})
$$
and expanding the logarithm in a power series in $\L ^2$
$$
\oint_{A_k}\d x{A'\over A}
=2\pi i\,\ak+\sum_{n=1}^{\infty}{(-1)^n\over
n}\L^{2n}\oint_{A_k}\d x  ({T(x)\over C(x)})^n
\eqno(3.6)
$$
These line integrals can be easily evaluated by the
method of residues, since the cycle $A_k$ is a contour
enclosing the pole $\ak$, and
we obtain at once
$$
\oint_{A_k}\d x x{A'\over A}=2\pi i\,\ak
+\sum_{n=1}^{\infty}{(-1)^n\over n}\L^{2n}
({\partial\over\partial \ak})^{n-1}R_k(\ak) ^n
\eqno(3.7)
$$
The second term on the right hand side of (3.3) is evaluated
using the identity
$$
x({A'\over A}-\12 {B'\over B})
({B\over
A^2})^m=-{\d\over \d x}\big[{x\over 2m}({B\over A^2})^m\big]
+{1\over 2m}({B\over A^2})^m
\eqno(3.8)
$$
and the following Taylor expansion in powers of $T(x)$ :
$$
{B^m(x)\over A^{2m}(x)}=\sum_{n=0}^{\infty}
{\G(2m+n)\over\G(2m)\G(n+1)}{B(x)^mT(x)^n\over C(x)^{2m+n}}
\bigl ( - { \L ^2 \over 4} \bigr ) ^n
$$
In terms of the functions $R_k(x)$ and $S_k(x)$, we readily get
$$
\oint_{A_k}\d x{B(x)^m\over A(x)^{2m}}
=2\pi i\sum_{n=0}^{\infty}{(-1)^n\L^{2m+2n}\over\G(2m)\G(n+1)}
\bigl ({\partial\over\partial \ak} \bigr )^{2m+n-1}
(R_k(\ak)^nS_k(\ak)^m)
\eqno(3.9)
$$
Assembling (3.7) and (3.9), and using the $\G$-function identity
$$
\G(\12)\G(2m+1)=2^{2m}\G(m+\12)\G(m+1)
$$
we obtain the following
expansion to all orders in $\L$ for the renormalized order parameter $a_k$
$$
a_k=\ak+\sum_{{m,n\geq0 \atop m+n>0}}
{(-1)^n\L^{2m+2n}\over (m!)^2n!2^{2m}}
\bigl ({\partial\over\partial \ak} \bigr )^{2m+n-1}(R_k(\ak)^nS_k(\ak)^m)
\eqno(3.10)
$$
As we already noted above, the special case $T(x)=0$ is of particular 
interest, since it suffices for us to determine
the prepotential in all generality. For later reference, we list
here the corresponding expansion for $a_k$ 
$$
a_k=\ak+\sum_{m=1}^{\infty}
{\L^{2m}\over 2^{2m}(m!)^2}
\bigl ({\partial\over\partial\ak} \bigr )^{2m-1}S_k(\ak) ^m.
\eqno(3.11)
$$
Notice that this expansion is analytic in powers of $\L ^2$, as expected
from general arguments.

\medskip

\noindent {\bf b) Expansion of the branch points $x_k^\pm$ in powers of $\L$}

We shall now solve the branch point equation $A(x_k ^\pm)^2 - B(x_k ^\pm) =0$
in a power series in $\L$, around a solution $\ak$ of $C(x)=0$. The problem is
conveniently reformulated in terms of the functions $R_k$ and $S_k$ introduced in
(3.4-5) :
$$
x_k^{\pm}=\ak\pm \L\,S(x_k^{\pm}) ^{\12} - \L ^2 R_k(x_k ^\pm)
\eqno(3.12)
$$
Since $S_k ^{\12}$ and $R_k$ are analytic for $\ak - \al$ and $\ak +m_j$
away from 0, it follows that $x_k ^\pm$ is an analytic function in $\L$,
and admits a Taylor series expansion of the form
$$
x_k^{\pm}=\ak+\sum_{m=1}^{\infty}(\pm)^m\L^m\delta_k^{(m)}
\eqno(3.13)
$$ 
We begin by deriving an explicit formula for $\delta _k ^{(m)}$ in the
special case where $R_k(x)=0$, i.e. $T(x)=0$, first.
Then, viewing $x_k^-$ as a function of $x(\L)$, and using the 
defining equation (3.12), we find
$$
\delta_k^{(m)}
={(-1)^m\over m!}{\partial ^m \over\partial \L ^m}  x(\L) \big \vert_{\L=0}
={(-1)^{m-1}\over (m-1)!}{\partial ^{m-1}\over\partial  \L ^{m-1}}S
_k(x(\L))^{\12} 
\big \vert_{\L=0}
$$
The last expression can be put under the form of a contour integral
$$
\delta_k^{(m)}={(-1)^{m-1}\over 2\pi i}\oint_{C_0}{\d\L\over\L^m}\big (S_k
x(\L)\big) ^\12
$$
where $C_0$ is a small contour in the $\L$ plane, enclosing the point
$\L =0$ once. Making the change of variables $\L\rightarrow x=x(\L)$ and
using $\L(x)=-(x-\ak)S_k (x)^{-\12}  $, the above contour integral becomes
$$
\delta_k^{(m)}=-{1\over 2\pi i}\oint_{C_{\ak}}\d x \biggl [
{(S_k(x)^{\12})'S_k (x)^{m-1\over2}\over (x-\ak)^{m-1}}
-{S_k(x)^{m\over 2}\over (x-\ak)^m}\biggr ]
$$
where $C_{\ak}$ is now a contour enclosing $x=\ak$ once. Integrating by parts
gives at once the following key formula for $\delta_k^{(m)}$
$$
\delta_k^{(m)}={1\over m!} \bigl ({\partial\over\partial\ak}
\bigr )^{m-1}S_k(\ak)^{m\over 2}
\eqno(3.14)
$$
The expression for the general case, where $R_k(x)\not=0$, 
is easily read off from
(3.14) by performing the substitution $S_k (x) ^\12 
\longrightarrow S_k (x) ^\12 \mp \L
R_k (x)$, which gives for (3.13)
$$
x_k^{\pm}=\ak+\sum_{m=1}^{\infty}
\L^m
{1\over m!} \bigl ({\partial\over\partial\ak}
\bigr )^{m-1} \bigl [ \pm S_k(\ak) ^\12 - \L R_k (\ak) \bigr ] ^m
\eqno(3.15)
$$ 
Identifying powers in $\L$ yields the coefficients 
$\delta _k ^{(m)}$ for general
non-zero $S_k$ and $R_k$
$$
\delta_k^{(m)}= \sum _{n=0} ^{m-1} 
{(-1)^n \over (m-2n)! n!} \bigl ({\partial\over\partial\ak}
\bigr )^{m-n-1} \bigl ( S_k(\ak)^{{m\over 2}-n} R_k(\ak) ^n \bigr )
\eqno(3.16)
$$
 
\medskip

\noindent {\bf c) Method of analytic continuation}

The complete determination of both the quantum order parameters $a_k$
and the branch points $x_k^\pm$ in terms of exact Taylor series 
expansions in $\L$ was possible with the use of residue calculations.
Clearly, it would be desirable to carry over these methods, as much as
possible, to the calculation of the dual order parameters $a_{D,k}$,
(and thus to the prepotential $\F$) given by
$$
2\pi i ~a_{D,k}
=\oint_{B_k}\d\lambda
=2\int_{\x1}^{\xk}\d x {x({A'\over A}-\12{B'\over B})
  \over\sqrt{1-{B\over A^2}}}
\eqno(3.17)
$$
(Note that $\sqrt{A^2}=-A$ for $x$ on the path from $x_1^-$ to $\xk$.)
Evaluation of these integrals is more delicate than for the A-periods,
since the end point $x_k^-$ of the integration is within a distance of
order $\L$ from the points $\ak$ and $a_k$. Thus, naively expanding
the denominator square root in a power series may run into convergence
problems.

This difficulty may be circumvented with the use of the following
analytic continuation prescription. We introduce an auxiliary complex
parameter $\xi$, with $|\xi|\leq 1$, and we define
$$
2\pi i ~a_{D,k} (\xi)
=\oint_{B_k}\d\lambda
=2\int_{\x1}^{\xk}\d x {x({A'\over A}-\12{B'\over B})
  \over\sqrt{1- \xi ^2 {B\over A^2}}}
\eqno(3.18)
$$
We consider this expression throughout the unit disc $|\xi | \leq 1$,
and use analyticity at the origin to expand the denominator square
root in a power series in $\xi ^2$.
$$
2\pi i\,a_{D,k} (\xi)
=2\sum_{m=0}^{\infty}{\G(m+\12)\over\G(\12)\G(m+1)}
\xi^{2m}\int_{\x1}^{\xk}\d x 
x({A'\over A}-\12{B'\over B})({B\over A^2})^m 
\eqno(3.19)
$$
The original quantity $a_{D,k}$ is then obtained by analytic
continuation of the function $a_{D,k}$ to $\xi =1$. (In practice, we
shall find that singularities may occur at $\xi =1$ at intermediate
stages of the calculations, but cancel out in the final form for
$a_{D,k}$; thus, the final analytic continuations are trivial.)
 
\medskip

\noindent {\bf d) Non-renormalization theorems}

We are now in a position to prove that the only logarithmic
contributions, both in the dynamical scale $\L$ and in the differences
$a_k-a_l$ are those generated by 1-loop perturbation theory, as given
by the second line in (2.1). As our starting point, we use the
expansion (3.19) for $a_{D,k}$, and further expand the denominators in
$A(x)$ in a power series in $T(x)$. As a result, each term in this
double series is now a rational function in $x$ : a polynomial in $B$
and in $T$, divided by a power of $C(x)$, and for $m=0$, a single term
of the form $B'/B$. Again, from residue calculus, we have a general
expression for the {\it residues at the simple poles} : they are given
by contour integrals around that pole
$$
\eqalign{
{x({A'\over A}-\12{B'\over B}) \over\sqrt{1- \xi ^2 {B\over A^2}}}
= &
\12 \sum _{j=1} ^{N_f} { m_j \over x+m_j}
+ \sum _{l=1} ^{N_c} {1 \over x-\al} {1\over 2 \pi i} \oint _{A_l} dy
{y({A'\over A}-\12{B'\over B}) \over\sqrt{1- \xi ^2 {B\over A^2}}}
\cr
&+ (N_c - \12 N_f) + 
\sum _{l=1} ^{N_c} \sum _{p=2} ^\infty {M_l ^{(p)} (\xi) \over
(x-\al)^p}
\cr}
\eqno (3.20)
$$
Letting $\xi \longrightarrow 1$, we recognize the residues of the
simple poles at $\al$ as the quantum order parameters $a_l$, so that
$$
\eqalign{
2 \pi i a_{D,k}
= &
  \sum _{j=1} ^{N_f}  m_j \l( x_k ^- +m_j)
+2 \sum _{l=1} ^{N_c} a_l \l(x_k^ - -\al)
\cr  
& +2 (N_c - \12 N_f)x_k^-   
- 2\sum _{l=1} ^{N_c} \sum _{p=2} ^\infty { 1 \over p-1} {M_l ^{(p)} (1)
\over (x_k ^- -\al)^{p-1}}
\cr}
\eqno (3.21)
$$
Here, as in the remainder of the paper,
we have written explicitly in the expression for
$2\pi i\,a_{D,k}$ only the contributions
of the upper integration bound $x_k^-$ in the
integral (3.19). Evidently, the lower
bound $x_1^-$ contributes a similar term
with $k$ replaced by $1$, and the opposite sign.
In particular, we observe that we are free to
add $k$-independent constants to (3.21), since
such constants will cancel between the two
bounds of integration.

Using the exact result for $x_k ^- -\ak$, from (3.12), and the fact
that $x_k ^-$ and $a_k$ are analytic functions of $\L$, with Taylor
series coefficients that are rational functions of the parameters $\ak$,
it is clear that
$$
\eqalignno{
2 \pi i a_{D,k}
= &
2 \l \L ~a_k
+ \sum _{j=1} ^{N_f} (a_k + m_j) \l (a_k +m_j)
- \sum _{l \not= k} ^{N_c} (a_k - a_l) \l (a_k - a_l )^2
\cr
&
+ {\rm \ power \ series \ in \  }\L {\rm \ with \ coefficients \ rational
\ in \ } a_l.
&(3.22) \cr}
$$
The first three terms on the right hand side arise from the 
1-loop contribution to
the effective prepotential.

To see this, we need to clarify the implications of the constraint
$\sum_{k=1}^{N_c}a_k=0$. 
The prepotential $\F$ of (2.1) is a function
$\F(a_1,\cdots,a_{N_c})$ of {\it all} variables $a_k$, whose
restriction to the
constraint hyperplane $\sum_{k=1}^{N_c}a_k=0$
is the prepotential appearing in (2.2). Written
in terms of $\F(a_1,\cdots, a_{N_c})$, the
condition (2.2) becomes
$$
a_{D,k}={\partial\over\partial a_k}\F(a_1,\cdots,a_{N_c})-
{\partial\over\partial a_1}\F(a_1,\cdots, a_{N_c}),\ 2\leq
k\leq N_c
\eqno(3.23)
$$
Conveniently, such differences occur automatically in expressions
of the form (3.17), if we take into account the contributions
of the lower integration bound $x_1^-$. Thus, it suffices
to identify the gradient of the prepotential function $\F$, with {\it all}
variables $a_1,\cdots, a_k$, viewed as independent.
Doing so, the first line of (3.22) is readily recognized as a gradient
$$
-{1\over 4}{\partial\over\partial a_k}[
\sum_{l,m=1}^{N_c}(a_l-a_m)^2\l
{(a_l-a_m)^2\over\Lambda^2}-\sum_{l=1}^{N_c}\sum_{j=1}^{N_f}(a_l+m_j)^2
\l{(a_l+m_j)^2\over\Lambda^2}]\eqno(3.24)
$$
up to power series in $\Lambda$ with coefficients rational
in $a_k$. These are exactly the logarithmic
singularities indicated in (2.1), due to perturbative
corrections.
As for 
the instanton contributions themselves, the
d-instanton contribution $\F_d$ is homogeneous of degree $d$ in the $a_i$'s,
and thus the Euler relation implies
$$
d\F_d=\sum_{l=1}^{N_c}a_l{\partial \F_d\over\partial a_l}
$$
Since ${\partial \F_d\over\partial a_l}$ is logarithm-free, so is $\F_d$.
This completes the proof of the non-renormalization theorem.

\medskip

As we worked it out above, the dual order parameters $a_{D,k}$ and thus the
prepotential $\F$ are a sum of the classical and 1-loop 
perturbative contributions,
plus a power series expansion in powers of $\L$. Actually, general
arguments suggest that the power series should involve only 
even powers of $\L$.
This fact is not completely obvious from the construction 
of $a_{D,k}$ in
(3.17) : while the integrand is even in $\L$, the integration 
limit $x_k^-$ is
not even. However, it is clear from the form of the integral 
in (3.17), that the
part of $a_{D,k}$ which is odd in $\Lambda $ is obtained by 
taking the
following difference 
$$
\eqalign{
2\pi i  ~\bigl (a_{D,k} (\L) -  a_{D,k} (- \L) \bigr )
=&
2  \int _{x_k ^+}^{x_k ^-} 
\d x {x({A'\over A}-\12{B'\over B}) \over\sqrt{1-{B\over A^2}}}
-2  \int _{x_1 ^+} ^{x_1 ^-} 
\d x {x({A'\over A}-\12{B'\over B}) \over\sqrt{1-{B\over A^2}}}
\cr
= & 
-2 \pi i (a_k - a_1)
\cr}
$$
This linear addition is merely a modification of the classical value of the
prepotential, and is innocuous. Thus, we see that the instanton contributions
are purely even in $\L$, as required by general arguments.

\medskip

\noindent {\bf e) Expansion of the dual order parameter $a_{D,k}$}

In this subsection, we shall work out an explicit formula for the dual order
parameters $a_{D,k}$, in a series expansion in powers of $\L$. In view of the
general arguments
\footnote{*}{We have checked the validity of these general arguments by 
explicit
calculation to 1- and 2-instanton orders; these calculations will not be
presented here.}
 advanced in sect. II.c, the dual order parameters
$a_{D,k}$ and the prepotential $\F$, when expressed as functions of the
renormalized order parameters $a_k$, are independent of the function $T(x)$.
Thus, henceforth, we shall set $T(x)=0$ without loss of generality. 

When $T(x)=0$, the function $A(x)$ simplifies and is given by 
$$A(x)=C(x).$$ 
Our
starting point for the calculation of the dual order parameters $a_{D,k}$ is
then the series expansion of (3.19), which for $T(x)=0$ becomes
$$
2\pi i\,a_{D,k} (\xi) 
=
2\sum_{m=0}^{\infty}{\G(m+\12)\over\G(\12)\G(m+1)} ~ \xi^{2m}
\int_{\x1}^{\xk}\d x ~
x({C'\over C}-\12{B'\over B})({B\over C^2})^m 
\eqno (3.25)
$$
The $m=0$ terms in  (3.25) can be easily integrated, and we find
$$
\eqalign{
2\int_{\x1}^{\xk}\d x ~x ({C'\over C}-\12 {B'\over B})
=
(2N_c-N_f)\xk
&
+2\sum_{l=1}^{N_c}\al\l(\xk-\al) 
\cr
&
+\sum_{m=1}^{N_f}m_j\l(\xk+m_j)
\cr}
\eqno(3.26)
$$
As for the remaining $m>0$ terms,
it is convenient to perform an initial integration by parts
using the identity (3.8), and rewrite them as
$$
\int_{\x1}^{\xk}\d x x({C'\over C}-\12{B'\over B})({B\over C^2})^m
=-{1\over 2m}\xk+{1\over 2m}\int_{x_1^-}^{\xk}\d x({B\over C^2})^m
\eqno(3.27)
$$
The contribution of the first term on the right hand side of (3.27)
can be summed explicitly, and we find (c.f. Appendix A) 
$$
\eqalignno{
2\pi i \,a_{D,k}
=&
(2N_c-N_f-2\l 2)\xk
+2\sum_{l=1}^{N_c}\al\l(\xk -\al)
+\sum_{j=1}^{N_f}m_j\l(\xk+m_j)\cr
&\qquad
+2 \sum_{m=1}^{\infty}{\G(m+\12)\over\G(\12)\G(m+1)}
{\xi^{2m}\over 2m}\int_{\x1}^{\xk}\d x \bigl ({B\over C^2}\bigr )^m
&(3.28)\cr}
$$
Next, we expand the rational function $B^m/C^{2m}$ into partial
fractions. Since $N_f<2N_c$, the function $B^m/C^{2m}$
vanishes as $x\rightarrow\infty$,
so that, with suitable coefficients
$Q_{l;p}^{(2m)}$, we have the expansion
$$
\bigl ({B\over C^2}\bigr )^m=\sum_{l=1}^{N_c}
\sum_{p=1}^{2m} \L ^{2m} Q_{l;p}^{(2m)}{1\over (x-\al)^p}
\eqno(3.29)
$$
We now make us of the observation  of sect. III.d on the structure of the
logarithmic terms : for $p=1$ in (3.29), the coefficient
$Q_{l;1}^{(2m)}$ is just the residue at the pole $\al$ of $B^m/A^{2m}$,
and can thus be expressed as a contour integral around $\al$, so that
$$
\L ^{2m} Q_{l;1}^{(2m)}
={2m\over 2\pi i}\oint_{A_l}\d x x({C'\over C}-\12{B'\over B})({
B\over C^2})^m
$$
The contributions of these terms to $a_{D,k}$ are now easily recognizable
$$
\eqalignno{
\sum_{m=1}^{\infty}{\G(m+\12)\over\G(\12)\G(m+1)}
{\xi^{2m}\L^{2m}\over 2m}Q_{l;1}^{(2m)}
&={1\over 2\pi i}\oint_{A_l}\d x x({C'\over C}-\12{B'\over B})({1\over
\sqrt{1-\xi^2{B\over C^2}}}-1)\cr
&=a_l-\al
&(3.30)\cr}
$$
Substituting in (3.28), and carrying out the $\d x$ integrations,
we arrive at the following formula, basic for our subsequent calculations
$$
\eqalignno{
2\pi i \,a_{D,k}=&(2N_c-N_f-2\l 2)\xk
+2\sum_{l=1}^{N_c}a_l\l (\xk-\al)+\sum_{j=1}^{N_f}m_j\l (\xk+m_j)\cr
&\qquad-
2
\sum_{m=1}^{\infty}{\G(m+\12)\over\G(\12)\G(m+1)}
{\xi^{2m}\over 2m}\sum_{l=1}^{N_c}\sum_{p=2}^{2m}
{Q_{l;p}^{(2m)}\over p-1}{\L^{2m}\over (\xk-\al)^{p-1}}
&(3.31)\cr}
$$
The coefficients $Q_{l;p} ^{(2m)}$ can be obtained from a Taylor expansion of
the function $S_k(x)$ combined with (3.4), and we obtain the simple 
expressions 
$$
Q_{l;p}^{(2m)}
={1\over (2m-p)!}\bigl ({\partial\over\partial\al}\bigr )^{2m-p}S_l(\al)^m
\eqno(3.32)
$$
It remains to express the quantities $x_l ^-$ and $\al$, as well as the
coefficients $Q_{l;p}^{(2m)}$ in terms of the renormalized order parameters
$a_l$. The relevant formulas were already derived in 
(3.11), (3.13) and (3.14).
Carrying out these substitutions in practice is rather cumbersome however, and
will be done in sect. IV for 1- and 2-instanton contributions, i.e. up to and
including corrections of order $\L ^4$.

\bigskip

\centerline{\bf IV. THE 1- and 2- INSTANTON CONTRIBUTIONS}

\bigskip

In this section, we derive explicit expressions for the  prepotential, 
including
perturbative (1-loop), 1- and 2- instanton contributions (i.e. up to and
including order $\L ^4$). As was argued in sect. II.c,  without loss of
generality, the prepotential can be evaluated  from the curve with $T(x)=0$,
which we assume from now on. Thus, our starting point is the expansion of 
(3.31), which we now wish to express in terms of $a_l$, up to and including
order $\L^4$. To avoid any confusion of orders of expansion, it is most
systematic (though perhaps somewhat lengthier) to first express all quantities
in terms of the parameters $\al$, and then convert the expression into a
function solely of the renormalized order parameters $a_l$.

\medskip

\noindent {\bf a) Expansion of the branch points $x_k ^-$}

Our calculations can be carried out either in terms of the derivatives at
$\ak$ of the functions $S_k(x)$, or, equivalently, in terms of the quantities
$\delta_k^{(m)}$, defined in (3.13) and expressed in terms of $S_k(x)$ in (3.14).
It turns out that the calculations can be presented most concisely in terms of
$\delta _k ^{(m)}$. 
From (3.13), we recall that, to order $\L ^4$, we have
$$
\eqalign{
x_k^- = & \ak 
- \L \delta _k ^{(1)} 
+ \L ^2 \delta _k ^{(2)}
-\L ^3 \delta _k ^{(3)}
+ \L ^4 \delta _k ^{(4)}
\cr
a_k = & \ak 
+ { 1\over 2} \L ^2 \delta _k ^{(2)} 
+ { 3 \over 8} \L ^4 \delta _k^{(4)}
\cr}
\eqno (4.1)
$$
Also, the expansion up to order $\L^3$ of powers of $\xk-\ak$
$$
\eqalign{
(\xk-\ak)^{-p}=
\biggl [ 
&1+\L p{\dkt\over\dko}+\L^2\bigl \{ -p{\dkT\over\dko}+\12 p(p+1)
({\dkt\over\dko})^2\bigr \} 
+\L^3 \bigl \{ p{\dkf\over \dko} \cr &
-p(p+1){\dkt\dkT\over(\dko)^2}
+{p(p+1)(p+2)\over 6}{\dkt\over \dko}\bigr \}
\biggr ](-1)^p\L^{-p}(\dko)^{-p}
\cr}
\eqno (4.2)
$$
as well as the expansion up to order $\L ^2$ of powers of $x_k -\al$, 
for $k\not=l$, 
$$
\eqalign{
(\xk-\al)^{-p}=(\ak - \al) ^{-p}
\biggl [
1
+\L p{\dko\over\ak-\al} &
+\L^2 \bigl \{ -p{\dkt\over\ak-\al} 
\cr
&
+\12 p(p+1){(\dko)^2\over(\ak-\al)^2} \bigr \}
\biggr ]
\cr}
\eqno (4.3)
$$
will be of particular use.

\medskip

\noindent{\bf b) Evaluation of the coefficients $Q_{l;p}^{(2m)}$}

A general expression for the residue functions $Q_{l;p}^{(2m)}$ in terms of
$S_k(x)$ was given in (3.32). In terms of the functions $\delta_k^{(m)}$, they
are easily worked out to low order as follows
$$
\eqalignno{
Q_{l;2m}^{(2m)}\ =&(\dlo)^{2m}\cr
Q_{l;2m-1}^{(2m)}=&2m\dlt(\dlo)^{2m-2}\cr
Q_{l;2m-2}^{(2m)}=&2m\dlT(\dlo)^{2m-3}+m(2m-3)(\dlt)^2(\dlo)^{2m-4}\cr
Q_{l;2m-3}^{(2m)}=&2m\dlf(\dlo)^{2m-4}+2m(2m-4)\dlt\dlT(\dlo)^{2m-5}\cr
&\quad+{1\over 3}m(2m-4)(2m-5)(\dlt)^3(\dlo)^{2m-6}
&(4.4)\cr}
$$

\medskip 

\noindent  {\bf c) Summation of the series in $m$}

It is convenient to separate the series in $m$ in (3.31)
into two terms (I) and (II), corresponding respectively
to the contributions of the poles at $\al$, $l\not=k$ and  at $\ak$.
$$
\eqalignno{
{\rm (I)}=&
2
\sum_{m=1}^{\infty}{\G(m+\12)\over\G(\12)\G(m+1)}
{\xi^{2m}\over 2m}\sum_{l\not=k}\sum_{p=2}^{2m}
{Q_{l;p}^{(2m)}\over p-1}{\L^{2m}\over (\xk-\al)^{p-1}}
\cr
{\rm (II)}=&
2
\sum_{m=1}^{\infty}{\G(m+\12)\over\G(\12)\G(m+1)}
{\xi^{2m}\over 2m}\sum_{p=2}^{2m}
{Q_{k;p}^{(2m)}\over p-1}{\L^{2m}\over (\xk-\ak)^{p-1}}
&(4.5)\cr}
$$
We concentrate first on (I). In this case $\xk-\al\sim \L ^0$,
the contributions up to and including terms of order
$\L^4$ arise only from $m\leq 2$, and so
we may set $\xi=1$ immediately. We obtain
$$
{\rm (I)}
=\sum_{l\not=k}\biggl [
 {1\over 2}\L^2{Q_{l;2}^{(2)}\over \xk-\al}
+{1\over 16}\L^4{Q_{l;4}^{(4)}\over (\xk-\al)^3}
+{3\over 32}\L^4{Q_{l;3}^{(4)}\over (\xk-\al)^2}
+{3\over 16}\L^4{Q_{l;2}^{(4)}\over \xk-\al}
\biggr ]
\eqno(4.6)
$$
We can now substitute in the expression (4.6) for $Q_{l;p}^{(2m)}$ in terms
of the $\delta_k^{(n)}$, and expand $(\xk-\al)^{-p}$ as in (4.3). The result
is
$$
\eqalignno{{\rm (I)}
=&\sum_{l\not=k}\biggl [
{1\over 2}\L^2{(\dlo)^2\over\ak-\al}
+{1\over 2}\L^3{(\dlo)^2\dko\over(\ak-\al)^2}
-{1\over 2}\L^4{(\dlo)^2\dkt\over(\ak-\al)^2}
+{1\over 2}\L^4{(\dlo)^2(\dko)^2\over(\ak-\al)^3}\cr
&\qquad+{1\over 16}\L^4{(\dlo)^4\over(\ak-\al)^3}
+{3\over 8}\L^4{(\dlo)^2\dlt\over(\ak-\al)^2}
+{3\over 4}\L^4{\dlo\dlT\over\ak-\al}
+{3\over 8}\L^4{(\dlt)^2\over\ak-\al} \biggr ]
&(4.7)\cr}
$$

\medskip

We turn next to contribution (II). Here $\L^{2m}(\xk-\ak)^{-p+1}\sim\L^{2m-p+1}$,
so that contributions to two-instanton order are received from all terms
satisfying $2m-3\leq p\leq 2m $.
(We note that, in particular, all $m$ contribute to any instanton order.) 
Thus,
we have
$$
\eqalignno{
{\rm (II)}
=& 2\sum_{m=1}^{\infty}{\G(m+\12)\over\G(\12)\G(m+1)}
{\xi^{2m}\over 2m}\L^{2m}\cr
&\quad\times\biggl [{Q_{k;2m}^{(2m)}\over(2m-1)(\xk-\ak)^{2m-1}}\theta _{m-1}
+{Q_{k;2m-1}^{(2m)}\over(2m-2)(\xk-\ak)^{2m-2}}\theta _{m-2}
\cr
&\quad+{Q_{k;2m-2}^{(2m)}\over(2m-3)(\xk-\ak)^{2m-3}}\theta _{m-2}
+{Q_{k;2m-3}^{(2m)}\over(2m-4)(\xk-\ak)^{2m-4}}\theta _{m-3}
 \biggr ]
&(4.8)\cr}
$$
Here, we have denoted by $\theta _m $ the Heaviside function defined by 
$\theta _m =1$ for $m \geq0$, and $\theta _m =0$ otherwise. 
We may now use the
values for $Q_{k;p}^{(2m)}$ found earlier (4.4),
and the expansions for $(\xk-\ak)^{-p}$ of (4.2) to arrive at
the following formula, valid to order $\L^4$ included.
$$
\eqalignno{
{\rm (II)}
=&
-2\L\dko\sum_{m=1}^{\infty}{\G(m+\12)\over\G(\12)\G(m+1)}{1\over 2m(2m-1)}
+
4\L^2\dkt\sum_{m=2}^{\infty}{\G(m+\12)\over\G(\12)\G(m+1)}{1\over 2m(2m-2)}\cr
&
-6\L^3\dkT\sum_{m=2}^{\infty}{\G(m+\12)\over\G(\12)\G(m+1)} {1\over 2m(2m-3)}
+
8\L^4\dkf\sum_{m=3}^{\infty}{G(m+\12)\over\G(\12)\G(m+1)}{1\over 2m(2m-4)}\cr
&
-{1\over 2}\L^2\dkt 
+{1\over 2}\L^3\dkT
-{11\over 16}\L^4\dkf
-{1\over 2}\L^3{(\dkt)^2\over\dko}
+{1\over 4}\L^4{\dkt\dkT\over\dko}
-{3\over 8}\L^4{(\dkt)^3\over(\dko)^2}
&(4.9)\cr}
$$
We note that all series in $m$ are now convergent, which
is why $\xi$ has simply been set to 1. The values of the series that occur in
(4.9) have been worked out in Appendix A. Inserting them in (4.9), we find
$$
\eqalignno{
{\rm (II)}=&\L\dko(2\l2-2)+\L^2\dkt({1\over 2}-\l 2)
+\L^3\big[\dkT(-{5\over 3}+2\l 2)-{1\over 2}{(\dkt)^2\over\dko}\big]\cr
&\quad+\L^4\big[\dkf({9\over 16}-{5\over 4}\l 2)
+{1\over 4}{\dkt\dkT\over\dko}-{3\over 8}{(\dkt)^3\over(\dko)^2}\big]
&(4.10)\cr}
$$

\medskip

\noindent {\bf d) Rearranging logarithmic contributions}

Our next task is to rewrite the logarithmic terms, which are
so far expressed in terms of $\xk-\al$, in terms of the
renormalized order parameters $a_k$'s. We shall do so
in two stages, first replacing $\xk$ by $\ak$, and then
replacing both $\ak$ and $\al$ by $a_k$ and $a_l$.
\medskip
It is convenient to exploit the fact
that $\xk$ satisfies the branch point equation (3.3),
and to rewrite the logarithm terms in (3.19) first as
$$
\eqalignno{
2\sum_{l=1}^{N_c}&a_l\l(\xk-\al)+\sum_{j=1}^{N_f}m_j\l(\xk+m_j)
-2(\l\L)\xk = {\rm -(III)-(IV)-(V)}
\cr
{\rm (III)}
=&2\sum_{l\not=k}^{N_c}(\xk-\al)\l(\xk-\al) 
-\sum_{j=1}^{N_f}(\xk+m_j)\l(\xk+m_j)
\cr
{\rm (IV)}
=&2\sum_{l\not=k}^{N_c}(\al-a_l)\l(\xk-\al)
\cr
{\rm (V)}
=&2(\xk-a_k)\l(\xk-\ak)
&(4.11)\cr}
$$
To evaluate (III), we expand the functions $\l(\xk-\al)$ and 
$\l(x_k + m_j)$ in 
power series around $\l(\ak-\al)$ and $\l (\ak +m_j)$ respectively.
We obtain this way
$$
\eqalignno{
{\rm (III)}=&(2N_c-2-N_f)(\xk-\ak)+2\sum_{l\not=k}^{N_c}(\ak-\al)
\l(\ak-\al)-\sum_{j=1}^{N_f}(\ak+m_j)\l(\ak+m_j)\cr
&-(\xk-\ak)\l S_k(\ak)-\12(\xk-\ak)^2{\partial\over\partial \ak}\l 
S_k(\ak)\cr
&-{1\over 6}(\xk-\ak)^3{\partial^2\over\partial \ak^2}\l S_k(\ak)
-{1\over 24}(\xk-\ak)^4{\partial^3\over\partial \ak^3}\l S_k(\ak)
&(4.12)\cr}
$$
As before, the derivatives of $S_k(\ak)$ can be readily
expressed in terms of the $\delta_k^{(q)}$
$$
\eqalignno{
{\partial\over\partial \ak}\l S_k(\ak)&=2{\dkt\over(\dko)^2}\cr
{\partial^2\over\partial \ak^2}\l S_k(\ak)&=4{\dkT\over(\dko)^3}
-6{(\dkt)^2
\over(\dko)^4}\cr
{\partial^3\over\partial \ak^3}\l S_k(\ak)&=
12{\dkf\over(\dko)^4}-48{\dkt\dkT\over(\dko)^5}+
40{(\dkt)^3\over(\dko)^6}
&(4.13)\cr}
$$
Expanding as well the powers of $\xk-\ak$ in $\L$,
we obtain the following expression for (III)
$$
\eqalignno{
{\rm (III)}=&
(2N_c-2-N_f)(\xk-\ak)+2\sum_{l\not=k}^{N_c}(\ak-\al)
\l(\ak-\al)-\sum_{j=1}^{N_f}(\ak+m_j)\l(\ak+m_j)\cr
&-2(\xk-\ak)\l\dko-\L^2\dkt+{2\over 3}\L^3\dkT+\L^3{(\dkt)^2\over\dko}\cr
&-\12\L^4\dkf+{1\over 3}\L^4{(\dkt)^3\over(\dko)^2}-2
\L^4{\dkt\dkT\over\dko}
&(4.14)\cr}
$$
Similarly, expansions for (IV) and (V) in terms of $\L$ are easily
determined using the expansions for $a_k$ and $\xk$ given in (3.12) 
and (4.1-2).
We find for (IV)
$$
\eqalignno{
{\rm (IV)}=&-\L^2\sum_{l\not=k}\dlt\l(\ak-\al)
-{3\over 4}\L^4\sum_{l\not=k}\dlf
\l(\ak-\al)\cr
&+\L^3\sum_{l\not=k}{\dko\dlt\over\ak-\al}
-\L^4\sum_{l\not=k}{\dlt\dkt\over\ak-\al}
+\12\L^4\sum_{l\not=k}{\dlt(\dko)^2\over(\ak-\al)^2}
&(4.15)\cr}
$$
and for (V)
$$
\eqalignno{{\rm (V)}=
&2(-\L\dko+\12\L^2\dkt-\L^3\dkT+{5\over 8}\L^4\dkf)( \l\L + \l \dko )\cr
&+2\L^2\dkt-2\L^3\dkT+2\L^4\dkf+\L^4{\dkt\dkT\over\dko}
+{1\over 6}\L^4{(\dkt)^3\over(\dko)^2}
&(4.16)\cr}
$$
For computational purposes, we had replaced the coefficients $a_k$
in front of the log terms in (3.19) by $\ak$. Upon assembling
the expressions we just obtained for (III)-(V), and
recognizing the combinations making up $a_k$ from $\ak$,
we obtain
$$
\eqalignno{
(-2\l{\L\over 2}&-2N_c+N_f)\xk
+2\sum_{l=1}^{N_c}(\xk-a_l)\l(\xk-\al)-
\sum_{j=1}^{N_f}(\xk+m_j)\l(\xk+m_j)\cr
=&
2\l2\,\xk-2\l\L ~a_k+(N_f-2N_c)\ak
-2(\xk-\ak)\cr
&
+2\sum_{l\not=k}(\ak-a_l)\l(\ak-\al)-
\sum_{j=1}^{N_f}(\ak+m_j)\l(\ak+m_j)\cr
&
+\L^2\dkt
-\L^2\dkt\l\dko
-{3\over 4}\L^4\dkf\l\dko
-{4\over 3}\L^3\dkT
+{3\over 2}\L^4\dkf\cr
&
+\L^3{(\dkt)^2 \over \dko}
+\12\L^4{(\dkt)^3\over(\dko)^2}
-\L^4{\dkt\dkT\over \dko}\cr
&
+\L^3\sum_{l\not=k}{\dko\dlt\over\ak-\al}
-\L^4\sum_{l\not=k}{\dkt\dlt\over\ak-\al}
+\12\L^4\sum_{l\not=k}{(\dko)^2\dlt\over(\ak-\al)^2}
&(4.17)\cr}
$$
In view of the identity
$$
-\L^2\dkt\l\dko-{3\over 4}\L^4\dkf\l\dko
=-2(\ak-a_k)\sum_{l\not=k}\l(\ak-\al)+(\ak-a_k)\sum_{j=1}^{N_f}
\l(\ak+m_j)
$$
the preceding expression simplifies to
$$
\eqalignno{
(-2\l{\L\over 2}&-2N_c+N_f)\xk
+2\sum_{l=1}^{N_c}(\xk-a_l)\l(\xk-\al)-
\sum_{j=1}^{N_f}(\xk+m_j)\l(\xk+m_j)\cr
=&
2\l 2\xk
-2\l\L ~ a_k+(N_f-2N_c)\ak
-2(\xk-\ak)\cr
&
+2\sum_{l\not=k}(a_k-a_l)\l(\ak-\al)
-\sum_{j=1}^{N_f}(a_k+m_j)\l(\ak+m_j)\cr
&
+\L^2\dkt
-{4\over 3}\L^3\dkT
+{3\over 2}\L^4\dkf
+\L^3{(\dkt)^2\over\dko}
+\12\L^4{(\dkt)^3\over(\dko)^2}
-\L^4{\dkt\dkT\over\dko}\cr
&
+\L^3\sum_{l\not=k}{\dko\dlt\over\ak-\al}
-\L^4\sum_{l\not=k}{\dkt\dlt\over\ak-\al}
+\12\L^4\sum_{l\not=k}{(\dko)^2\dlt\over(\ak-\al)^2}
&(4.18)\cr}
$$
We now replace $\ak$ by $a_k$, using the expansion (3.12).
The logarithms expand as
$$
\eqalignno{
2\sum_{l\not=k}&(a_k-a_l)\l(\ak-\al)
-\sum_{j=1}^{N_f}(a_k+m_j)\l(\ak+m_j)\cr
=&2\sum_{l\not=k}(a_k-a_l)\l(a_k-a_l)
-\sum_{j=1}^{N_f}(a_k+m_j)\l(a_k+m_j)\cr
&\quad-\12\L^2\dkt(2N_c-N_f)-{3\over 8}\L^4\dkf(2N_c-N_f)\cr
&\quad-{1\over 4}\L^4\sum_{l\not=k}{(\dkt-\dlt)^2\over a_k-a_l}
+{1\over 8}\L^4\sum_{j=1}^{N_f}{(\dkt)^2\over a_k+m_j}
&(4.19)\cr}
$$
Altogether, we have
$$
\eqalignno{(-2&\l{\L\over 2}-2N_c+N_f)\xk
+2\sum_{l=1}^{N_c}(\xk-a_l)\l(\xk-\al)-
\sum_{j=1}^{N_f}(\xk+m_j)\l(\xk+m_j)\cr
=&2(\l 2)\xk+(-2\l\L+N_f-2N_c)a_k\cr
&+2\sum_{l\not=k}(a_k-a_l)\l(a_k-a_l)-
\sum_{j=1}^{N_f}(a_k+m_j)\l(a_k+m_j)\cr
&+2\L\dko-\L^2\dkt+{2\over 3}\L^3\dkT-\12\L^4\dkf
+{3\over 4}\L^4{(\dkt)^3\over(\dko)^2}+\L^3{(\dkt)^2\over\dko}-
\L^4{\dkt\dkT\over\dko}\cr
&-\12\L^4\sum_{l\not=k}{\dkt\dlt\over\ak-\al}
-{1\over 4}\L^4\sum_{l\not=k}{(\dlt)^2\over\ak-\al} 
+\L^3\sum_{l\not=k}{\dko\dlt\over\ak-\al} 
+\12\L^4\sum_{l\not=k}{(\dko)^2\dlt\over\ak-\al}
&(4.20)\cr}  
$$

\medskip

\noindent {\bf e) Assembling all contributions}

We are now ready to assemble all the components 
(4.7), (4.10) and (4.20) of
$a_{D,k}$.  Several cancellations take place in 
view of the first group
of resummation identities in Appendix B, and we obtain
$$
\eqalignno{
2\pi i \,a_{D,k}=&-(2\l 2-2\l\L+N_f-2N_c)a_k
\cr
&
-2\sum_{l\not=k}(a_k-a_l)\l(a_k-a_l)+\sum_{j=1}^{N_f}(a_k+m_j)\l(a_k+m_j)\cr
&
+\12\L^2\dkt-\12\L^2\sum_{l\not=k}{(\dlo)^2\over\ak-\al}
-{1\over 8}\L^4\sum_{l\not=k}{(\dlt)^2\over\ak-\al}\cr
&
+{7\over 16}\L^4\dkf-{1\over 8}\L^4{(\dkt)^3\over(\dko)^2}+
{1\over 4}\L^4{\dkt\dkT\over\dko}\cr
&
+{1\over 4}\L^4\sum_{l\not=k}{\dkt(\dlo)^2\over(\ak-\al)^2}-
{1\over 16}\L^4\sum_{l\not=k}{(\dlo)^4\over(\ak-\al)^3}\cr
&
-{3\over 8}\L^4\sum_{l\not=k}{(\dlo)^2\dlt\over(\ak-\al)^2}-
{3\over 4}\L^4\sum_{l\not=k}{\dlo\dlT\over\ak-\al}
&(4.21)\cr}
$$
Our penultimate task, before we are in position to
identify the prepotential, is to eliminate the
$\al$'s, and rewrite the whole
expression for $a_{D,k}$ entirely in terms of the renormalized
$a_l$'s. We observe in the above that the terms $\delta_l^{(p)}$
are written in terms of the $\bar a_m$'s. We temporarily stress this feature
by denoting them by $\delta_l^{(p)}(\bar a)$, while letting
the simpler notation $\delta_l^{(p)}$ stand rather for their
counterparts $\delta_l^{(p)}(a)$. The passage from
$\delta_l^{(p)}(\bar a)$ to $\delta_l^{(p)}(a)\equiv\delta_l^{(p)}$ 
is easily worked out
for $p=1,2$, using the explicit expressions (B.10) (B.11) for 
$\delta_l^{(p)}$
$$
\eqalignno{
(\dlo(\bar a))^2
=&(\dlo)^2-\L^2(\dlt)^2-\L^2\sum_{m\not=l}
{\dmt(\dlo)^2\over a_l-a_m} & (4.22)
\cr
\dlt(\bar a)
=&\dlt +\12\L^2{(\dlt)^3\over(\dlo)^2}
-\L^2{\dlt\dlT\over\dlo}
-\L^2\sum_{m\not=l}{\dlt\dmt\over a_l-a_m}
+\12\L^2\sum_{m\not=l}
{(\dlo)^2\dmt\over(a_l- a_m)^2}\cr}
$$
Applying (4.22), we can now express $a_{D,k}$ in terms of the
$\delta_l^{(p)}(a) $'s:
$$
\eqalignno{
2 \pi i \,a_{D,k}
=&(2\l \L-2\l2+2N_c-N_f)a_k 
& (4.23) \cr
&
-2\sum_{l\not=k}(a_k-a_l)\l(a_k-a_l)+\sum_{j=1}^{N_f}
(a_k+m_j)\l(a_k+m_j)\cr
&
-\12\L^2\sum_{l\not=k}{(\dlo)^2\over a_k-a_l}+\12\L^2\dlt
+{1\over 8}\L^4{(\dkt)^3\over(\dko)^2}+{7\over 16}\L^4\dkf
-{1\over 4}\L^4{\dkt\dkT\over\dko}
\cr
&
+{3\over 8}\L^4\sum_{l\not=k}{(\dlt)^2\over a_k-a_l}
-{1\over 8}\L^4\sum_{l\not=k}{(\dlo)^2\dlt\over(a_k-a_l)^2}
+\12\L^4\sum_{l\not=k}\sum_{m\not=l}{\dmt(\dlo)^2\over(a_l-a_m)(a_k-a_l)}\cr
&
-\12\L^4\sum_{l\not=k}{\dkt\dlt\over a_k-a_l}
+{1\over 4}
\sum_{l\not=k}{(\dko)^2\dlt\over(a_k-a_l)^2}
-{1\over 16}\L^4\sum_{l\not=k}{(\dlo)^4\over(a_k-a_l)^3}
-{3\over 4}\L^4\sum_{l\not=k}{\dlo\dlt\over a_k-a_l}\cr}
$$
As was emphasized after equation (3.21), we recall
that (4.23) incorporates implicitly a similar, opposite,
contribution with the index $k$ replaced by the index 1.

\medskip

\noindent  {\bf f) Integrability conditions and the prepotential}

We show now how to use the summation identities in Appendix B
and rearrange the terms so as to exhibit the prepotential.
The fact that this can at all be done provides a powerful
check on the final answer. As noted already before,
(c.f. (3.22)) the classical part together with perturbative 
corrections $\F^{(0)}$ is
easy to identify
$$
\eqalignno{\F^{(0)}=&
{1\over 2\pi i}(2N_c-N_f-\l 2)\sum_{l=1}^{N_c}a_k^2 & (4.24) \cr
&-{1\over 8\pi i}\big(\sum_{l,m=1}^{N_c}(a_l-a_m)^2\l
{(a_l-a_m)^2\over\Lambda^2}-\sum_{l=1}^{N_c}\sum_{j=1}^{N_f}(a_l+m_j)^2
\l{(a_l+m_j)^2\over\Lambda^2}\big)
\cr}
$$
Next, it follows immediately from
the definition of $\dkt$ and the summation identity
(B.3) that the one-instanton contribution
to the prepotential is given by
$$
\F^{(1)}={1\over 8 \pi i}\L^2\sum_{l=1}^{N_c}(\dlo)^2
\eqno(4.25)
$$
We shall identify the two-instanton potential
$\F^{(2)}$ in several steps. First, we use the identity (B.8)
to rewrite the corresponding terms $a_{D,k}^{(2)}$ as
$$
\eqalignno{
2 \pi i \,a_{D,k}^{(2)}
=&{1\over 16}\L^4
{\partial\over\partial a_k}\biggl [
\sum_{l\not=k}\sum_{m\not=l,k}{(\dmo)^2(\dlo)^2\over (a_l-a_m)^2}
\biggr ]
-{1\over 4}\L^4\sum_{l\not=k}{(\dko)^2(\dlo)^2\over(a_k-a_l)^3}
\cr 
&
-{1\over 4}\L^4\sum_{l\not=k}{\dlo\dlT\over a_k-a_l}
+{1\over 8}\L^4\sum_{l\not=k}{(\dlt)^2\over a_k-a_l}
-{1\over 8}\L^4\sum_{l\not=k}{(\dlo)^2\dlt\over(a_k-a_l)^2}\cr
&-{1\over 2}\L^4\sum_{l\not=k}{\dkt\dlt\over a_k-a_l}
+{1\over 4}\L^4\sum_{l\not=k}{(\dko)^2\dlt\over(a_k-a_l)^2}
-{1\over 16}\L^4\sum_{l\not=k}{(\dlo)^4\over(a_k-a_l)^3}\cr
&+{1\over 8}\L^4{(\dkt)^3\over(\dko)^2}
+{7\over 16}\L^4\dkf
-{1\over 4}\L^4{\dkt\dkT\over\dko}
&(4.26)\cr}
$$
Use now the identities (B.5) and (B.6) to exhibit the terms involving
$\dlT$ and $(\dlt)^2$ in terms of a derivative with respect to $a_k$. We find
$$
\eqalignno{
2 \pi i \,a_{D,k}^{(2)}=&
{1\over 16}\L^4{\partial\over\partial a_k}
\big[\sum_{l\not=k}\sum_{m\not=l,k}{(\dmo)^2(\dlo)^2\over (a_l-a_m)^2}
+\sum_{l\not=k}\big(\dlo\dlT-\12(\dlt)^2\big)\big]\cr
&-{1\over 4}\L^4\sum_{l\not=k}{(\dko)^2(\dlo)^2\over(a_k-a_l)^3}
+{1\over 4}\L^4\sum_{l\not=k}{(\dko)^2\dlt\over(a_k-a_l)^2}
-{1\over 2}\L^4\sum_{l\not=k}{\dkt\dlt\over a_k-a_l}\cr
&+{1\over 8}\L^4{(\dkt)^3\over(\dko)^2}
+{7\over 16}\L^4\dkf
-{1\over 4}\L^4{\dkt\dkT\over\dko}
&(4.27)\cr}
$$
The identities (B.1) and (B.2) now apply to produce
$$
\eqalignno{
2 \pi i \,a_{D,k}^{(2)}=&
{1\over 16}\L^4
{\partial\over\partial a_k}\big[
\sum_{l\not=k}\sum_{m\not=l,k}{(\dmo)^2(\dlo)^2\over (a_l-a_m)^2}
+\sum_{l\not=k}\big(\dlo\dlT-\12(\dlt)^2\big)\big]\cr
&-\12\L^4\sum_{l\not=k}{(\dko)^2(\dlo)^2\over (a_k-a_l)^3}
+{1\over 4}
\L^4\sum_{l\not=k}{\dkt(\dlo)^2\over (a_k-a_l)^2}\cr
&+{1\over 8}\L^4{(\dkt)^3\over(\dko)^2}
-{1\over 4}\L^4{\dkt\dkT\over \dko}
+{3\over 16}\L^4\dkf
&(4.28)\cr}
$$
The terms in the second row of (4.28) are just the ones occurring
in the identity (B.7). As for the remaining local terms, they
are also integrable, in view of the following simple
local identities, which are a direct consequence
of their definition (4.2)
$$
\dkf={1\over 6}{\partial\over\partial a_k}[(\dkt)^2+2\dko\dkT]
\eqno(4.29)
$$
$$
{\dkt\dkT\over\dko}-\12{(\dkt)^3\over(\dko)^2}={1\over 4}
{\partial\over\partial a_k}(\dkt)^2
\eqno(4.30)
$$
We can now write the two-instanton contribution to $a_{D,k}$
under the form
$$
2 \pi i ~a_{D,k}^{(2)}={1\over 16}
\L^4
{\partial\over\partial a_k}\bigg[
\sum_{l\not=m}{(\dmo)^2(\dlo)^2\over (a_l-a_m)^2}
+\sum_{l=1}^{N_c}(\dlo\dlT-\12(\dlt)^2)\bigg]
\eqno(4.31)
$$

\medskip

\noindent {\bf g) Summary of result for the prepotential up to two 
instanton contributions}

We summarize here the main formulas, written in terms of the functions
$$
S_k(x)={\prod_{j=1}^{N_f}(x+m_j)\over
\prod_{l\not=k}(x-a_l)^2}
\eqno(4.32)
$$ 
The dual order parameters $a_{D,k}$'s
are given by
$$
a_{D,k}={\partial\over\partial a_k}\F(a)
$$
with
$$
\F=\F^{(0)}+\F^{(1)}+\F^{(2)}+O(\Lambda^{3(2N_c-N_f)})
$$
and
$$
\eqalignno{
2\pi i\F^{(0)}=&-{1\over 4}
\sum_{l\not=k}(a_k-a_l)^2\l{(a_k-a_l)^2\over\Lambda^2}
+{1\over 4}\sum_{j=1}^{N_f}(a_k+m_j)^2\l{(a_k+m_j)^2\over\Lambda^2}\cr
&-(\l 2-2N_c+ N_f)\sum_{k=1}^{N_c}a_k^2 
& (4.33a) \cr
2\pi i\F^{(1)}=&{1\over 4}\Lambda^{2N_c-N_f}\sum_{l=1}^{N_c}S_l(a_l)
& (4.33b) \cr
2\pi i\F^{(2)}=&{1\over 16}\Lambda^{2(2N_c-N_f)}\biggl [ ~
\sum_{l\not=m}{S_l(a_l)S_m(a_m)\over (a_l-a_m)^2}
+{1\over 4}\sum_{l=1}^{N_c}S_l(a_l){\d^2S_l\over\d a_l^2}(a_l) ~\biggr ]
&(4.33c)\cr}
$$

The $\l 2$ terms can be eliminated by a redefinition of the
scale ${\Lambda^{N_c-\12 N_f}\over 2}\rightarrow \Lambda^{N_c-\12 N_f}$,
upon which the prepotential takes the more familiar form
$$
\eqalignno{
2\pi i\F^{(0)}=&-{1\over 4}
\sum_{l\not=k}(a_k-a_l)^2\l{(a_k-a_l)^2\over\Lambda^2}
+{1\over 4}\sum_{j=1}^{N_f}(a_k+m_j)^2\l{(a_k+m_j)^2\over\Lambda^2}\cr
&+(2N_c- N_f)\sum_{k=1}^{N_c}a_k^2 
& (4.34a) \cr
2\pi i\F^{(1)}=&\Lambda^{2N_c-N_f}\sum_{l=1}^{N_c}S_l(a_l)
& (4.34b) \cr
2\pi i\F^{(2)}=&\Lambda^{2(2N_c-N_f)}\biggl [ ~
\sum_{l\not=m}{S_l(a_l)S_m(a_m)\over (a_l-a_m)^2}
+{1\over 4}\sum_{l=1}^{N_c}S_l(a_l){\d^2S_l\over\d a_l^2}(a_l) ~\biggr ]
&(4.34c)\cr}
$$
By construction, these contributions to the effective prepotential are
invariant under permutations of the variables $a_k$, and hence are invariant
under the Weyl group of SU($N_c$). It is of course possible, though in general
very cumbersome,  to rewrite these results in terms of symmetric 
polynomials in
the $a_k$, such as $s_i$ and $\sigma _i$, defined in (2.4) and (2.5). However,
the above results are perfectly well-defined and invariant as they stand.

\bigskip

\centerline{\bf V. SPECIAL CASES AND DISCUSSION}

\bigskip

In this section, we evaluate some of our results in various special cases
discussed in the literature either directly from the quantum field theory 
point
of view, using instanton calculations or from the Seiberg-Witten type 
approach.

\medskip

\noindent {\bf a) Comparisons with quantum field theory instanton 
calculations}

Results directly from instanton calculations in quantum field theory are
available as follows. For the simplest case of two colors, $N_c=2$, and
$N_f<2N_c$, various tests of the exact results of [1] were performed
in [8,14], while 2-instanton results were obtained in [15], and are found
to be in agreement with [1]. For the case of general number of colors,
$N_c$, results appears to be available only for contributions involving just a
single instanton [7], and are found to agree with the exact results of [2]. 

\medskip

\noindent {\bf b) $N_c=2$ results}

A useful check on the correctness of the coefficients in (4.33) 
is provided by a
comparison with the exact results for $N_c=2$ and $N_f=0$. 
We set $\bar a = \bar
a_1 = - \bar a_2$, $a=a_1=-a_2$, $\bar \Lambda = \Lambda ^2$ 
and we find from the
definition of $S_k$ that
$$
\eqalign{
S_1(a_1) = &
S_2(a_2) = {1 \over 4 a^2}, 
\cr
{\partial ^2 S_1(a_1) \over \partial a_1 ^2} 
= &{\partial ^2 S_1(a_1) \over \partial a_1 ^2} 
={3 \over 8 a^6}
\cr}
\eqno (5.1)
$$
Using these results, we find the effective prepotential up to 2 instanton
corrections
$$
\F = { i \over 2\pi} \bigl [ 2a^2 \l 4a^2 + (2 \l 2 -4 \l \Lambda -4) a^2
-{\Lambda ^4 \over 8 a^2 } - {5 \Lambda ^8 \over 2^{10} a^6} +{\cal
O}(\Lambda ^{12}) \bigr ]
\eqno (5.2)
$$
This result agrees with the expansion of the exact results of [1] in powers of
$\Lambda$, as calculated in [6], provided we make the redefinition $\Lambda ^2
\longrightarrow \Lambda ^2 /2$.

\medskip

It is not much more difficult to incorporate the effects of $N_f$ matter
hypermultiplets. We shall do this here for $N_f=3$; the result for lower 
values
of $N_f$ may then be obtained by decoupling. We thus find for $N_f=3$ 
$$
\eqalignno{
2 \pi  i \F ^{(1)} 
= &
	{\Lambda _3 \over 8 a^2} \bigl [ a^2(m_1+m_2+m_3) +m_1m_2m_3 \bigr ] 
&(5.3)\cr
2 \pi  i \F ^{(2)} 
= &
	{\Lambda _3 ^2 \over 2^{10} a^6} 
		\bigl [ a^6 + a^4 (m_1 ^2 + m_2 ^2 + m_3 ^2 ) 
			- 3a^2 (m_1^2 m_2 ^2 + m_2 ^2 m_3 ^2 + m_3 ^2 m_1^2)
			+5 m_1 ^2 m_2 ^2 m_3 ^2 \bigr ]
\cr}
$$
Decoupling the third hypermultiplet, by letting $\Lambda _3 m_3 = \Lambda _2
^2$, and sending $m_3 \to \infty$, we obtain for $N_f=2$
$$
\eqalign{
2 \pi  i \F ^{(1)} 
= &
	{\Lambda _2 ^2 \over 8 a^2} \bigl [ a^2 +m_1m_2 \bigr ] 
\cr
2 \pi  i \F ^{(2)} 
= &
	{\Lambda _2 ^4 \over 2^{10} a^6} 
		\bigl [ a^4 - 3a^2 (m_1^2  + m_2 ^2 ) +5 m_1 ^2 m_2 ^2 \bigr ]
\cr}
\eqno (5.4)
$$
and decoupling also the second hypermultiplet, letting 
$\Lambda _2 m_2 = \Lambda
_1 ^2$ as $m_2 \to \infty$, we find for $N_f=1$
$$
\eqalign{
2 \pi  i \F ^{(1)} 
= &
	{\Lambda _1 ^3 \over 8 a^2} m_1 
\cr
2 \pi  i \F ^{(2)} 
= &
	{\Lambda _1 ^6 \over 2^{10} a^6} 
		\bigl [ - 3a^2 +5 m_1 ^2  \bigr ]
\cr}
\eqno (5.5)
$$
From the decoupling of the first hypermultiplet, with 
$\Lambda _1 m_1 = \Lambda
^2$, as $m_1 \to \infty$, we recover the result of (5.2).

\medskip

We notice that for $N_c=2$, the expression for the renormalized 
order parameter
$a$ in terms of the classical value $\bar a$ is readily worked 
out to all orders
with the help of the expression for $\delta _1 ^{(m)}$
$$
\delta _1 ^{(m)} = (-) ^m {\G (m- \12) \over \G (-\12) \G (m+1) } 
{1 \over \bar
a ^{2m-1}}
\eqno (5.3)
$$
which yields, using (3.11)
$$
a = \bar a \sum _{m=0} ^\infty { \G (2m-\12) \over 2^{2m} \G (-\12)  \G (m+1) ^2}
\biggl ( {\Lambda \over \bar a } \biggr ) ^{4m}
\eqno(5.4)
$$
This sum is easily recognized as the hypergeometric function $\bar a
F(-{1\over 4},{1\over 4};1; {\Lambda ^4 \over \bar a^4})$, which is the
correct expression, as obtained in [6].

\medskip

\noindent {\bf b) $N_c=3$ results}

We can similarly work out the contributions to the prepotential from 1- and
2-instanton effects for $N_c=3$. We have $a_1+a_2+a_3=0$, and it is
customary to express the results in terms of the symmetric polynomials, 
$u$ and
$v$ and the discriminant $\Delta$
$$
u= -a_1a_2 -a_2a_3 -a_3a_1,
\qquad 
v=a_1 a_2 a_3;
\qquad \qquad
\Delta = 4u^3 -27v^3
\eqno (5.5)
$$
The functions $S_k(a_k)$ are easily evaluated and we have
$$
S_k(a_k) 
= {\prod _{j=1} ^{N_f} (a_k + m_j) \over \prod _{l\not=k} (a_k -a_l)} 
\eqno (5.6)
$$
Using (4.33), we find the one instanton results 
$$
\eqalign{
N_f = 5 \qquad & \qquad
2 \pi i \F ^{(1)} 
=
{\Lambda _5 \over 4 \Delta } \bigl [ - u^2 v + (2u^3 -9v^2) t_1 -3 uv t_2
+ 2 u^2 t_3 -9  v t_4 + 6u t_5 \bigr ]
 \cr
N_f = 4 \qquad & \qquad
2 \pi i \F ^{(1)} 
=
{\Lambda _4 ^2 \over 4 \Delta } \bigl [  (2u^3 -9v^2)  -3 uv t_1
+ 2 u^2 t_2 -9  v t_3 + 6u t_4 \bigr ]
\cr
N_f = 3 \qquad & \qquad
2 \pi i \F ^{(1)} 
=
{\Lambda _3 ^3 \over 4 \Delta } \bigl [  -3 uv 
+ 2 u^2 t_1 -9  v t_2 + 6u t_3 \bigr ]
\cr
N_f = 2 \qquad & \qquad
2 \pi i \F ^{(1)} 
=
{\Lambda _2 ^4 \over 4 \Delta } \bigl [  2 u^2  -9  v t_1 + 6u t_2 \bigr ]
\cr
N_f = 1 \qquad & \qquad
2 \pi i \F ^{(1)} 
=
{\Lambda _1 ^5 \over 4 \Delta } \bigl [  -9  v  + 6u t_1 \bigr ]
\cr
N_f = 0 \qquad & \qquad
2 \pi i \F ^{(1)} 
=
{6 u \Lambda ^6 \over 4 \Delta}
\cr}
\eqno(5.7)
$$
For the case $N_f=0$, these results are in agreement with those of [6], 
derived using Picard-Fuchs equations.

Two instanton effects may similarly be evaluated. For $N_f=0$, we find that
$$
2 \pi i \F ^{(2)} = {9 \Lambda ^{12} \over 16 \Delta ^3} ( 17 u^3 +189 v^2)
\eqno(5.8)
$$
in agreement with [6]. It is possible to express also the two instanton
corrections for $N_f >0$ in terms of the the symmetric polynomials 
$u$, $v$ and
$t_p$; the expressions become quite cumbersome however, and it is much better
to leave the results in the original forms of (4.33).

\bigskip
\bigskip

\centerline{\bf ACKNOWLEDGEMENTS}

\bigskip

One of us (E. D.) would like to thank the CERN theory divison and the Aspen
Center for physics for their hospitality while most of this work was being
carried out.

\bigskip

\vfill\break

\centerline{\bf APPENDIX A: SOME USEFUL NUMERICAL SERIES}

\bigskip

In this appendix, we derive the values of the series which are needed
in the evaluation of the dual periods $a_{D,k}$'s. Consider the series
$$
f_n^{(q)}=\sum_{m=q}^{\infty}{\G(m+\12)\over\G(\12)\G(m+1)}{1\over 2m-n}
\eqno(A.1)
$$
where $q$ and $n$ are some fixed integers satisfying $2q>n\geq 0$. 
Then we have
$$
\eqalignno{
f_0^{(1)}&=\l 2,\ f_0^{(2)}=\l 2-{1\over 4}\cr
f_1^{(1)}&=1,\ f_1^{(2)}=\12\cr
f_2^{(2)}&=\12\l 2+{1\over 4}\cr
f_3^{(2)}&={5\over 6}\cr
f_4^{(3)}&={3\over 8}\l 2+{9\over 32}&(A.2)\cr}
$$
To see this, we note that
$f_n^{(q)}$ is the value at $\xi=1$ of the corresponding function
$$
f_n^{(q)}(\xi)=
\sum_{m=q}^{\infty}{\G(m+\12)\over\G(\12)\G(m+1)}{1\over 2m-n} \xi ^{2m}
$$
which is easily seen to admit an integral representation. In fact
$$
\xi^{n+1}{\d\over\d\xi} \bigl (\xi^{-n}f_n^{(q)}(\xi)\bigr )
=\sum_{m=q}^{\infty}
{\G(m+\12)\over\G(\12)\G(m+1)}\xi^{2m}
={1\over\sqrt{1-\xi^2}}-\sum_{m=0}^{q-1}
{\G(m+\12)\over\G(\12)\G(m+1)}\xi^{2m}
$$
It follows that
$$
f_n^{(q)}(\xi)=\xi^n\int_0^{\xi}{\d\eta\over\eta^{n+1}}
\biggl [
{1\over\sqrt{1-\eta^2}}-\sum_{m=0}^{q-1}
{\G(m+\12)\over\G(\12)\G(m+1)}\eta^{2m}
\biggr ]
\eqno(A.3)
$$
which is an absolutely convergent integral for
$2q>n$. These integrals for $\xi=1$
can be evaluated explicitly by a change of variables to $\eta=2t(1+t^2)^{-1}$.
We find, e.g.,
$$
f_n^{(2)}(1)=2^{-n+1}\int_0^1\d t\, t^{-n+3}(t^2+3)(1+t^2)^{n-3}
$$
which leads easily to the values indicated in (A.2) for
$f_0^{(2)}$, $f_1^{(2)}$, $f_2^{(2)}$, and $f_3^{(2)}$
(and hence $f_0^{(1)}$ and $f_1^{(1)}$). For $f_4^{(3)}$,
it is somewhat faster to make the change of variables
$\eta=\sqrt{1-u^2}$, which leads to the integral of a
rational fraction
$$
f_4^{(3)}(1)=
\int_0^1{\d u\over (1-u^2)^3}\biggl [ 1-u-\12 u(1-u^2)-{3\over 8}
(1-u^2)^2\biggr ]
$$
This works out to the value given in (A.2).

\vfill\break

\centerline{\bf APPENDIX B: SUMMATION IDENTITIES}

\bigskip

The identities we need concern the sums over $l\not=k$
of expressions involving $(a_k-a_l)^{-p}$, and are
of two types. The first leads to local terms involving
only the $\delta_k^{(n)}$'s, while the second
leads to total derivatives with respect to $a_k$
of expressions which ultimately make up the prepotential.
The identities of the first type of interest to
us are the following
$$
\sum_{l\not=k}\big[{(\dlo)^2\over(a_k-a_l)^2}
+2{\dlt\over a_k-a_l}\big]=2{\dkT\over\dko}-({\dkt\over\dko})^2\eqno(B.1)
$$
$$
\sum_{l\not=k}\big[{(\dlo)^2\over(a_k-a_l)^3}
+{\dlt\over (a_k-a_l)^2}\big]
=-{\dkf\over(\dko)^2}+
2{\dkt\dkT\over(\dko)^3}-{(\dkt)^3\over(\dko)^4}
\eqno(B.2)
$$
To show these identities, we observe that the
pole expansion for ${B(x)\over C^2(x)}$ can be
rewritten as
$$
\sum_{l\not=k}\biggl [
{S_l(\al)\over(x-\al)^2}+{S_l'(\al)\over x-\al} \biggr ]
={1\over (x-\ak)^2}\big(S_k(x)-S_k(\ak)- S_k'(\ak)(x-\ak)\big)
$$
Letting $x\rightarrow\ak$ gives (B.1). Taking a higher order Taylor
expansion for $S_k(x)$ and letting again $x\rightarrow\ak$
after differentiation gives (B.2).
 
The identities of the second type depend on
the explicit forms for $\delta_k^{(q)}$. They are
$$
\eqalignno{
\sum_{l\not=k}{(\dlo)^2\over a_k-a_l}
=&-{1\over 2}{\partial\over\partial a_k}
\sum_{l\not=k}(\dlo)^2&(B.3)\cr
\sum_{l\not=k}{(\dlo)^4\over(a_k-a_l)^3}
=&-{1\over 6}{\partial\over\partial a_k}
\sum_{l\not=k}{(\dlo)^4\over(a_k-a_l)^2}
&(B.4)\cr}
$$
$$
\eqalignno{
\sum_{l\not=k} \biggl [
{(\dlo)^2\dlt\over(a_k-a_l)^2}
+2{(\dlt)^2\over a_k-a_l}\biggr ]
=&-\12{\partial\over\partial a_k}
\sum_{l\not=k}(\dlt)^2
&(B.5)\cr
\sum_{l\not=k}\big[{(\dlo)^4\over(a_k-a_l)^3}
+3{(\dlo)^2\dlt\over(a_k-a_l)^2}
+4{\dlo\dlT\over a_k-a_l}]=&
-{\partial\over\partial a_k}\sum_{l\not=k}\dlo\dlT&(B.6)\cr}
$$
$$
\sum_{l\not=k}\big[2{(\dlo)^2\dkt\over(a_k-a_l)^2}
-4{(\dlo)^2(\dko)^2\over(a_k-a_l)^3}\big]
={\partial\over\partial a_k}
\sum_{l\not=k}{(\dlo)^2(\dko)^2\over(a_k-a_l)^2}\eqno(B.7)
$$
$$
\eqalignno{\sum_{l\not=k}\sum_{m\not=l}{\dmt(\dlo)^2\over(a_l-a_m)(a_k-a_m)}
=&{1\over 8}{\partial\over\partial a_k}
\sum_{l\not=k}\sum_{m\not=l,k}{(\dmo)^2(\dlo)^2\over (a_l-a_m)^2}
-\12\sum_{l\not=k}{(\dlo)^2(\dko)^2\over(a_k-a_l)^3}\cr
&+\sum_{l\not=k}{\dlo\dlT\over a_k-a_l}
-\12\sum_{l\not=k}{(\dlo)^2\over a_k-a_l}&(B.8)\cr}
$$
We begin by establishing (B.4). The left hand side of the identity
can be rewritten
as
$$
\sum_{l\not=k}{1\over (a_k-a_l)^3}{\prod_{j=1}^{N_f}(a_l+m_j) ^2 \over
\prod_{m\not=l}(a_l-a_m)^4}
=\sum_{l\not=k}{1\over (a_k-a_l)^7}{\prod_{j=1}^{N_f}(a_l+m_j) ^2\over
\prod_{m\not=l,k}(a_l-a_m)^4}
$$
This is the same as
$$
-{1\over 6}{\partial\over\partial a_k}\sum_{l\not=k}
\sum_{l\not=k}{1\over (a_k-a_l)^6}{\prod_{j=1}^{N_f}(a_l+m_j) ^2\over
\prod_{m\not=l,k}(a_l-a_m)^4}
=-{1\over 6}\sum_{l\not=k}{1\over (a_k-a_l)^2}(\dlo)^4,
$$
as claimed. The identity (B.3) is proved
in the same way. Next, we note that for $l\not=k$
$$
{\partial\over\partial a_k}\delta_l^{(p)}
={1\over (p-1)!}{\partial^{p-1}\over\partial a_l^{p-1}}
\biggl \{{1\over a_l-a_k}(\dlo)^p \biggr \}
\eqno(B.9)
$$
For $p=1$, this follows by direct calculation from
the explicit formula for $\dlo$
$$
\dlo={1\over a_l-a_k}{\prod_{j=1}^{N_f}(a_l+m_j) ^{\12}
\over \prod_{m\not=l,k}(a_l-a_m)}
\eqno(B.10)
$$
The case of general $p$ follows next
from the definition (35) for $\delta_l^{(p)}$, which
expresses $\delta_l^{(p)}$ in terms of $\dlo$.
The identities (B.5), (B.6), and (B.7) are now easy to establish, simply
by differentiating the right hand side, and
applying (B.9).

Finally, we consider (B.8). We may use first the identity to
eliminate $\dmt$ in favor of $\dmo$
$$
\sum_{l\not=k}\sum_{m\not=l}{\dmt(\dlo)^2\over(a_l-a_m)(a_k-a_l)}
=-\12\sum_{l\not=k}\sum_{m\not=l}
{(\dmo)^2(\dlo)^2\over(a_l-a_m)^2(a_k-a_l)}
+\sum_{l\not=k}{\dlo\dlT\over a_k-a_l}
-\12\sum_{l\not=k}{(\dlt)^2\over a_k-a_l}\eqno(B.11)
$$
The double sum in the right hand side can be rewritten as
$$
\sum_{l\not=k}\sum_{m\not=l}
{(\dmo)^2(\dlo)^2\over(a_l-a_m)^2(a_k-a_l)}
=\sum_{l\not=k}{(\dko)^2(\dlo)^2\over (a_k-a_l)^3}
+\sum_{l\not=k}\sum_{m\not=l,k}
{(\dmo)^2(\dlo)^2\over(a_l-a_m)^2(a_k-a_l)}
$$
The second term above can now be recognized as a total derivative
$$
\sum_{l\not=k}\sum_{m\not=l}
{(\dmo)^2(\dlo)^2\over(a_l-a_m)^2(a_k-a_l)}
=-{1\over 4}{\partial\over\partial a_k}
\sum_{l\not=k}\sum_{m\not=l}
{(\dmo)^2(\dlo)^2\over(a_l-a_m)^2}
$$
where we made use of the symmetry in $l$ and $m$ of all the expressions
involved. The identity (B.8) follows.
\vfill\break

\centerline{\bf REFERENCES}
\bigskip
\item{[1]} N. Seiberg and E. Witten, Nucl. Phys. B 426 (1994) 19, hep-th/9407087;
Nucl. Phys. B 431 (1994) 484, hep-th/9408099.
\item{[2]} A. Klemm, W. Lerche, S. Yankielowicz, and S. Theisen,
Phys. Lett. B 344 (1995) 169;\hfil\break
P.C. Argyres and A. Faraggi, Phys. Rev. Lett. 73 (1995) 3931, hep-th/9411057;
\hfil\break
M.R. Douglas and S. Shenker, Nucl. Phys. B 447 (1995) 271, hep-th/9503163;
\hfil\break
P.C. Argyres, R. Plesser, and A. Shapere, Phys. Rev. Lett. 75 (1995) 1699,
hep-th/9505100;\hfil\break
J. Minahan and D. Nemeshansky, hep-th/9507032;\hfil\break
U.H. Danielsson and B. Sundborg, Phys. Lett. B 358 (1995) 273,
USITP-95-12, UUITP-20/95;
hep-th/9504102;\hfil\break
A. Brandhuber and K. Landsteiner, Phys. Lett. B 358 (1995) 73, hep-th/9507008;
\hfil\break
M. Alishahiha, F. Ardalan, and F. Mansouri, hep-th/9512005;\hfil\break
I. Pesando, hep-th/9506139.
\item{[3]} A. Hanany and Y. Oz, Nucl. Phys. B 452 (1995) 73, hep-th/9505075
\hfil\break
A. Hanany, hep-th/9509176.
\item{[4]} I.M. Krichever and D.H. Phong, hep-th/9604199, to
appear in J. of Differential Geometry;
\item{[5]} D. Amati, K. Konishi, Y. Meurice, G.C. Rossi, and G. Veneziano,
Phys. Rep. 162 (1988) 169.
\item{[6]} A. Klemm, W. Lerche, and S. Theisen, hep-th/9505150;\hfil\break
K. Ito and S.K. Yang, hep-th/96
\item{[7]} K. Ito and N. Sasakura, SLAC-PUB-KEK-TH-470, hep-th/9602073;
\item{[8]} N. Dorey, V. Khoze, and M. Mattis, hep-th/9606199, 
hep-th/9607202;\hfil\break
Y. Ohta, hep-th/9604051, hep-th/9604059;
\item{[9]} S. Kachru, A. Klemm, W. Lerche, P. Mayr, and C. Vafa,
Nucl. Phys. B 459 (1996) 537, hep-th/9508155;\hfil\break
A. Klemm, W. Lerche, P. Mayr, C. Vafa, and N. Warner, hep-th/9604034;
\item{[10]} M. Matone, Phys. Lett. B 357 (1995) 342;\hfil\break
G. Bonelli and M. Matone, Phys. Rev. Lett. 76 (1996) 4107;\hfil\break
G. Bonelli and M. Matone, hep-th/9605090;\hfil\break
T. Eguchi and S.K. Yang, Mod. Phys. Lett. A 11 (1996) 131;
\item{[11]} R. Donagi and E. Witten, hep-th/9511101;
\item{[12]} A. Gorsky, I.M. Krichever, A. Marshakov, A. Mironov, 
and A. Morozov,
Phys. Lett. B 355 (1995) 466, hep-th/9505035;\hfil\break
H. Itoyama and A. Morozov, hep-th/9512161, hep-th/9901168;\hfil\break
A. Marshakov, hep-th/9602005\hfil\break 
E. Martinec and N. Warner, hep-th/9509161, hep-th/9511052;\hfil\break
E. Martinec, hep-th/9510204,\hfil\break
C. Anh and S. Nam, hep-th/9603028,\hfil\break
T. Nakatsu and K. Takasaki, hep-th/9509162;
\item{[13]} I.M. Krichever, Comm. Pure Appl. Math. 47 (1994) 437.
\item{[14]} D. Finnell and P. Pouliot, Nucl. Phys. {\bf B453} (1995) 225;
\hfil\break
A. Yung, hep-th/9605096; \hfil\break
F. Fucito and G. Travaglini, hep-th/9605215
\item{[15]} T. Harano and M. Sato, hep-th/9608060

\end